\documentclass[sigconf]{acmart}

\AtBeginDocument{%
  \providecommand\BibTeX{{%
    \normalfont B\kern-0.5em{\scshape i\kern-0.25em b}\kern-0.8em\TeX}}}

\copyrightyear{2023}
\acmYear{2023}
\setcopyright{acmlicensed}\acmConference[CHIWORK 2023]{Annual Symposium on Human-Computer Interaction for Work 2023}{June 13--16, 2023}{Oldenburg, Germany}
\acmBooktitle{Annual Symposium on Human-Computer Interaction for Work 2023 (CHIWORK 2023), June 13--16, 2023, Oldenburg, Germany}
\acmPrice{15.00}
\acmDOI{10.1145/3596671.3597650}
\acmISBN{979-8-4007-0807-7/23/06}

\begin{document}

\title[The Impact of Artificial Intelligence on the Creativity of Knowledge Work]{Exploring Perspectives on the Impact of Artificial Intelligence on the Creativity of Knowledge Work: Beyond Mechanised Plagiarism and Stochastic Parrots}


\author{Advait Sarkar}
\affiliation{%
  \institution{Microsoft Research, University of Cambridge, University College London}
  \country{United Kingdom}}
\email{advait@microsoft.com}

\begin{abstract}
Artificial Intelligence (AI), and in particular generative models, are transformative tools for knowledge work. They problematise notions of creativity, originality, plagiarism, the attribution of credit, and copyright ownership. Critics of generative models emphasise the reliance on large amounts of training data, and view the output of these models as no more than randomised plagiarism, remix, or collage of the source data. On these grounds many have argued for stronger regulations on the deployment, use, and attribution of the output of these models.

However, these issues are not new or unique to artificial intelligence. In this position paper, using examples from literary criticism, the history of art, and copyright law, I show how creativity and originality resist definition as a notatable or information-theoretic property of an object, and instead can be seen as the property of a process, an author, or a viewer. Further alternative views hold that all creative work is essentially reuse (mostly without attribution), or that randomness itself can be creative. I suggest that creativity is ultimately defined by communities of creators and receivers, and the deemed sources of creativity in a workflow often depend on which parts of the workflow can be automated.


Using examples from recent studies of AI in creative knowledge work, I suggest that AI shifts knowledge work from material production to critical integration.  This position paper aims to begin a conversation around a more nuanced approach to the problems of creativity and credit assignment for generative models, one which more fully recognises the importance of the creative and curatorial voice of the users of these models, and moves away from simpler notational or information-theoretic views.
\end{abstract}




\received{20 February 2007}
\received[revised]{12 March 2009}
\received[accepted]{5 June 2009}

\maketitle

\section{Introduction}

In October 2022, creators using AI-generated art faced fierce opposition, including death threats, from the manga community in Japan \cite{deck_2022}, because the art strongly mimicked the styles of well-known manga artists, clearly enabled by the use of copyrighted illustrations in the training data. The Japanese anime and manga industries have long been accepting of fan art that directly reuses copyrighted work, and to some extent even encourages it, as it stimulates engagement with the franchise. But using AI to generate this artwork seemed to cross a line. It seemed to push beyond the limit of cultural acceptability. Something about the nature of mechanical production signals a shift in our relationship with creativity and knowledge work.

To begin to unpick this shift, we need a clearer articulation of creativity. This we can get from Margaret Boden \cite{boden_2007}, for whom a creative idea is surprising, novel, and valuable.\footnote{Variations of this ``three-criterion'' definition for creativity are widely adopted, as for example by the US Patent Office \cite{simonton2012taking}.} An idea may be novel with respect to the individual person (``P-creativity'') or with respect to a historical community (``H-creativity''). It is either the production of a novel (and surprising, and valuable) idea from within a pre-defined conceptual space, or the transformation of conceptual space to enable new forms of idea.

Blackwell develops this definition by observing that ``surprise'' can be defined in information-theoretic terms \cite{blackwell2022coding}. The information content of a message is a measure of how surprising it is; a message that is not surprising at all provides no new information. Moreover, drawing on Dennett's notion of the intentional stance \cite{dennett1971intentional}, Blackwell argues that even when AI produces something surprising, novel, and valuable, it cannot be creative, because that would incorrectly assume that the model itself has beliefs, goals, and desires. In other words, it cannot be creative because it is not \emph{intentional}. It would thus be incorrect to attribute creativity to the model, as it would be incorrect to attribute the capacity for arithmetic to a calculator \cite{collins1998shape}. Rather, any creativity in this instance arises from the human activity of constructing a ``performance'' on the basis of this random output, as a Tarot card reader produces a creative performance on the basis of a randomised arrangement of cards.

Moreover, Blackwell argues that the stochastic reuse of training data, even when arranged into novel forms, is not a truly novel idea; it is the execution and replay of human behaviour encoded into these systems. It is a form of objectivity laundering \cite{blackwell2020objective}, where inherently subjective human judgements are replayed in an algorithmically objectified manner. Yet these ``stochastic parrots'' \cite{bender2021dangers}, as Gebru, Bender, et al. term them, cannot be extricated from the subjective biases inherent in their training data. With this view of AI as mechanical replay of subjective judgements, Sarkar argues that the phrase ``human-AI collaboration'' ought to be avoided, because it implies that AI is a creative collaborator, responsible for a certain share of labour, rather than those who provided the training data, who are the true collaborators in any human-AI ``collaboration'' \cite{sarkar2023enough}.

With these precepts in place, it is possible to assess the creativity or plagiaristic nature of AI with respect to its training data and with respect to the wider world of information content. When an AI system repeats a sentence (or image, etc.) from its training data, it is not creative because it does not fulfil the criterion for new information, and it is a plagiarist because it can do so without attribution. When it repeats an idea or ideology from its training data, albeit in a unique and novel way, it is not creative because that idea derives from behaviour exemplified in the training set, and it may still be a plagiarist if it is the intellectual content of the idea and not the words used to describe it (or image used to depict it, etc.) that we consider to be original intellectual property \cite{lee2022language}. And when it produces a genuinely novel, surprising, and valuable idea, it may not be a plagiarist, but it cannot be considered creative because to do so would be to take an intentional stance.

There is extensive commentary on the nature of computers and creativity, far too much to include in this brief introduction (indeed practically every paper published in the long-running ACM conference on Creativity \& Cognition\footnote{\url{https://cc.acm.org/}} is a candidate for inclusion). There are relevant reviews of the field \cite{anantrasirichai2022artificial, boden2008computers, frich2018twenty, wang2017literature}. However I hope to have captured the salient points germane to the discussion of the current generation of artificial intelligence, primarily foundation models and large language models. These have been articulated by Boden, Blackwell, Bender, Dennett, Gebru, Sarkar, etc. as follows: creativity is information-theoretic surprise, with authorial intent. AI systems generate content through randomised reuse without intent or attribution; they are therefore not creative and tend to facilitate plagiarism. The danger is clear: integrating these systems into everyday knowledge workflows (writing documents, creating graphics, etc.) is likely to lead to a collective loss of creativity, an increase in mechanised plagiarism, and ``undermining creative economies'' \cite{weidinger2021ethical}.

These are valid, important, and logical perspectives on creativity and plagiarism. My aim in this paper is not to dismantle or refute them. However, I am concerned that these perspectives only represent a small fraction of the plurality of perspectives on creativity and plagiarism, all of which are equally valid, important, and logical. In this paper, I will show with examples the following alternative perspectives:

\begin{itemize}
    \item Process as creativity: that creativity lies not in the object of creation, but in the method of production (Section~\ref{sec:process}).
    \item Authorial intent and discourse as creativity: that the creative content of an object can only be understood with respect to the author's intent, or the societal discourse surrounding it (Section~\ref{sec:authorial_intent}).
    \item Interpretation as creativity: that creativity is not a property of an object but of the interaction between a reader and the object (Section~\ref{sec:interpretation}).
    \item Reuse as creativity: that reuse of prior material without attribution is an unavoidable aspect of any creative endeavour (Section~\ref{sec:reuse}).
    \item Randomness as creativity: that random processes can be viewed as creative of their own accord, especially as a mechanism to distance the (human) creator from the creation (Section~\ref{sec:randomness}).
\end{itemize}

After discussing these perspectives, I will explain how the form-content distinction becomes a recurrent issue in creativity (Section~\ref{sec:form_content}), and the challenges of relying on intellectual property law to help understand whether AI is being creative (Section~\ref{sec:intellectual_property}).

Each perspective raises several questions and harbours its own contradictions and ethical shortcomings. Answering these questions and resolving these contradictions is impossible; it is precisely the impossibility of this task that has given rise to such a plurality of perspectives. However impossible, the exercise is not futile. The result of considering these perspectives will be a new agenda for understanding how creativity in knowledge work is affected by AI, namely, as a shift from the direct production of knowledge artefacts to the \emph{critical integration} of AI output as part of a broader knowledge workflow (Section~\ref{sec:AI_natures}).

To clarify the scope of this paper, it is worth briefly discussing what is meant by AI, what is meant by knowledge work, and what is the relationship between creativity and knowledge work. First, ``AI''. AI is a broad term that encompasses an unusually diverse, heterogenous, and multidisciplinary set of concepts, stretching back over two centuries \cite{mackenzie2017machine}. When the term ``AI'' is used in this paper, it is referring to contemporary, statistical, deep learning approaches which use vast numbers of parameters and quantities of training data to model a space of knowledge artefacts (e.g., images, texts) and can be used to generate artefacts by sampling from this space. These include foundation models \cite{bommasani2021opportunities}, large language models \cite{brown2020language}, image generation models \cite{ramesh2021zero}, among others. These are often referred to collectively as ``generative AI''.

Next, ``knowledge work''. This term was popularised by Peter Drucker \cite{drucker1959}, for whom knowledge work was differentiated from manual work by its focus on applying mental faculties and knowledge acquired through systematic education. In this paper, I further refine the scope of the term by adopting the key property identified by Kidd \cite{kidd1994marks}, namely that knowledge work requires the private transformation of the individual doing the work, as an outcome of processing information. I further focus on those particular aspects of knowledge workflows that culminate in the production of material artefacts such as documents, textual communication, images, presentations etc., upon which there is a cultural or corporate expectation of creativity (as per Boden's criteria). These production-oriented tasks are directly affected by the AI technologies under discussion.

Finally, how does creativity relate to knowledge work? In some sense, the need for creativity in knowledge work falls out implicitly from the joint definitions of Boden and Kidd: in order to be usefully transformed by information processing, a knowledge worker must synthesize a novel idea. This process of synthesis is depicted by Wallas' multi-stage model of creativity \cite{wallas1926art}. Not all aspects of all knowledge work might be considered creative, and a notion of creativity may rather be dependent on individual and group norms, as Xu et al. argue \cite{xu2023creativity}. In this paper, rather than attempt to draw examples from the academic discourse on creativity in knowledge work, we will look at the discourse on creativity in literature, music, and art (among other fields). These are forms of knowledge work in which the concerns around creativity are acutely and deeply rooted, and consequently the discussion of creativity is particularly rich and nuanced. Each section will conclude with a reflection on how this discussion relates to the issues of AI and creativity in knowledge work more broadly.


\section{Alternative conceptions of creativity}
The current discourse around AI creativity and plagiarism focuses on the information content of its output. In particular, its output is measured and qualified in terms of familiar units of information: tokens, words, bits, or pixels. This information content is considered the primary or only determinant of the creativity of the output. 

However, there are alternative views of creativity which de-emphasise, or even ignore, the actual content of any objects produced as part of a creative process. These views move away from \emph{content} and towards \emph{context}. Sometimes, it is the process itself which is considered creative, and the resultant objects are incidental. Sometimes, an object can be viewed as creative or un-creative depending on the authorial intent and the social discourse which accompanies it, or the way in which it is interpreted by the viewer. Creativity may not be defeated by reuse, but entirely dependent on it. Random processes may themselves be sources of creativity. This section explains each of these views in turn.

\subsection{Process as creativity}
\label{sec:process}

The artistic practice of conceptual art emphasises the process of creating an artwork, and de-emphasises the resultant artefact \cite{lewitt1967paragraphs}. In conceptual art, according to LeWitt, \emph{``all of the planning and decisions are made beforehand and the execution is a perfunctory affair. The idea becomes a machine that makes the art''}. Conceptual artists proceed by experimentation, sketching, and extensive prototyping, until arriving at a method for executing or expressing an idea. This method can be executed once, or many times, but it is itself the true outcome of the creative process, not any particular execution. Seeking a complete separation between process and outcome, some practitioners proceeded \emph{``by making plans for their work and having these plans executed by others [...] not simply by describing a desired image, but by specifying the process by which the work was to be made''} \cite{galenson2011old}.

The ideology of conceptual art was prefigured in abstract expressionism, in particular ``action painters'' like Pollock and Gorky, who emphasised the bodily motion and physicality of the art process as essential to the content of the artwork. According to Froman \cite{froman1988action}, action painting shifts \emph{``from painting where [...] the artist paints toward the goal of an image [...] to painting where the artist comes to the canvas as a site for acting, so that the painting displays the event that takes place when the artist paints''}. It was anticipated perhaps earlier still in Brutalist architecture, which sought to expose and exalt materials and construction processes, producing \emph{``structures which were entirely visible, reducing historical ornament to a minimum so that the exterior could reflect the inner structure rather than hiding it''} \cite{clement2018brutalism}.


%

Conceptual art has a cognate practice in literature: conceptual writing \cite{acker2011against}. Conceptual writers proceed by writing with constraints (deriving from the OuLiPo movement \cite{james2009constraining}), or under a set of generative rules. Many work with computer programs, or write programs to facilitate the mechanised execution of writing rules (as in generative art \cite{boden2009generative}). The processes and rules are anterior to the generated texts, and are considered the objects in which creativity resides.



Computing gives us a ready metaphor for this view. We are already accustomed to thinking of algorithms as distinct creative contributions from any particular ``run'' of the algorithm with a certain input and output. Quicksort is a creative contribution distinct from mergesort or bubblesort, despite the fact that every correct sorting algorithm produces the same output as any other.

In a review of generative art practices, Boden and Edmonds point out that it is difficult to pinpoint the ``locus of creativity'', whether it is in the individual artwork, or in the ``art system'' \cite{boden2009generative}. This is not a question that can be logically answered from first principles, but is continuously renegotiated by communities of art producers and consumers. If in a particular context, creativity is viewed as a property of the process and not the outcome, whether AI is being creative or a plagiarist rests not on its output but rather whether its algorithm, or the human-AI-data complex which comprises the creational ``algorithm'', can be understood as distinct from or identical to another.

\subsection{Authorial intent and discourse as creativity}
\label{sec:authorial_intent}

In 1917 Marcel Duchamp signed a porcelain urinal and submitted it for an exhibition of the Society of Independent Artists. In doing so, he is credited with creating the single most influential work of modern art \cite{higgins_2004}. Though \emph{Fountain} was not Duchamp's first work repurposing ordinary materials, and others such as Picasso were experimenting with similar ideas at the time, its brazen statement of counterculture brought a sharp and unprecedented clarity to the underlying principle: Duchamp's ``readymades'' contended that art, and creativity, can come about from literally any object, through the pure force of authorial intent. The more general term for this kind of work is ``found art'', both in the sense that it is art that can be ``found'' in the environment, but also that it becomes art because it is ``found [to be] art'' through discourse. Duchamp explained that \emph{``an ordinary object [could be] elevated to the dignity of a work of art by the mere choice of an artist''} \cite{learning_2023}.

Duchamp's urinal may seem far removed from our everyday experience of creative authorship. Yet as Goldsmith notes \cite{acker2011against}, the vast majority of cultural transactions on the Internet revolve around readymades: \emph{``To be the originator of something that becomes a broader meme trumps being the originator of the actual trigger event that is being reproduced. The ``re-'' gestures -- such as reblogging and retweeting -- have become cultural rites of cachet in and of themselves. If you can filter through the mass of information and pass it on as an arbiter to others, you gain an enormous amount of cultural capital''}. In the era of social media, the curatorial intent behind re-sharing has come to carry its own cultural cachet of creativity. Such is the role of the ``trail blazer'' in Bush's world of the Memex \cite{bush1945we}, \emph{``who find delight in the task of establishing useful trails through the enormous mass of the common record''}. 


Authorial intent can do more than directly imbue an object with the property of being ``creative''. Often, the author's context and stated intent are used to disambiguate vocabulary and references, and to identify and interpret metaphors. Consider the following lines from Maya Angelou's \emph{Caged Bird} \cite{angelou1983caged}:

\begin{quote}
The free bird thinks of another breeze\\
and the trade winds soft through the sighing trees\\
and the fat worms waiting on a dawn bright lawn\\
and he names the sky his own.\\
\\
But a caged bird stands on the grave of dreams\\
his shadow shouts on a nightmare scream\\
his wings are clipped and his feet are tied\\
so he opens his throat to sing.
\end{quote}

Had Angelou been a well-known animal rights activist, the common interpretation of this poem would have been as an agitation against keeping birds in cages. But she was, of course, a civil rights activist, and with this knowledge one is justified in the interpretation that the birds are people, and the cage is racism. Works of art and literature generally do not include biographies of the authors and descriptions of their intent within the body of the work. It is context from outside the content of the work which is nevertheless essential for understanding the information content of the work, and evaluating it as creative or otherwise. Borges' delightful short story \emph{Pierre Menard, Author of the Quixote} \cite{borges1962ficciones} takes this idea to a logical extreme, in which the author Menard has written (though not copied) text identical to chapters from Cervantes' \emph{Don Quixote}, but which the narrator views as far superior due to differences in historical circumstance between Menard and Cervantes: \emph{``The Cervantes text and the Menard text are verbally identical, but the second is almost infinitely richer. [...] It is a revelation to compare the Don Quixote of Pierre Menard with that of Miguel de Cervantes''}.


As Foucault notes, access to as-yet unpublished notes or new information about an author, or even a return to familiar texts, can affect and revise our interpretation of works and their meaning \cite{foucault2017author}. A notable example of this is the publication in the late 1920s and early 1930s of Marx's \emph{Economic and Philosophic Manuscripts of 1844}, which nearly a century after their writing finally enabled a clearer interpretation of key concepts in Marxian philosophy, such as ``alienation'', which was poorly understood in Marx's own lifetime \cite{musto2010revisiting}.

Creative objects exist not as individual quanta of information, but in a reactive and shifting network. In this view, the meaning of an AI-generated text and therefore its status as creative, is not static, but may change depending on other texts that the AI generates, and other discourse that we draw upon as interpretive resources that support the text.

At this point, one might ask: to what extent is authorial intent relevant to AI? We do not want to attribute undue personhood or agency to the AI system itself (Dennett's intentional stance). Perhaps it is easier to argue that the human-AI-data complex responsible for a particular AI output is the author. However, it is also possible to argue that AI itself can be considered an author, without granting it personhood or agency.

Foucault considers the question of what an author entity, or ``author function'' actually is \cite{foucault2017author}, especially when used as an interpretive resource to help establish the meaning of a text: \emph{``Modern criticism, in its desire to `recover' the author from a work [... is] reminiscent of Christian exegesis [... in determining the authors of a text when it is unclear who wrote it.] According to Saint Jerome, there are four criteria: the texts that must be eliminated from the list of works attributed to a single author are those inferior to the others (thus, the author is defined as a standard level of quality); those whose ideas conflict with the doctrine expressed in the others (here the author is defined as a certain field of conceptual or theoretical coherence); those written in a different style and containing words and phrases not ordinarily found in the other works (the author is seen as a stylistic uniformity); and those referring to events or historical figures subsequent to the death of the author (the author is thus a definite historical figure in which a series of events converge)''}.

These criteria: a level of quality, conceptual coherence, stylistic uniformity, and historicity, can be applied to AI to give it an author function without also granting it agency or personhood. These ways of producing an author entity may seem imprecise in our modern world, where it is far easier to definitively establish the author of a text, and often possible to ask them directly: ``did you write this?'' and ``what did you mean by this?'' Even with these powers, unavailable to Saint Jerome, we still conceive of authors in Jeromian terms. For instance, Wittgenstein's later work contrasts so heavily with his earlier work that references often qualify him as ``early Wittgenstein'' or ``late Wittgenstein'', as though they were two separate people (this is also sometimes done with Marx to distinguish ``young Marx'' \cite{musto2015young}, presumably from ``old Marx'' but the latter is rarely used). Artists' lives are commonly periodised (e.g., Picasso's ``blue period''). This schizophrenic segmentation shows that we are willing to detach a singular person from the authorial entity or entities they gave rise to over the course of their life. Legal conceptions of authorship have similarly long been divorced of personhood, where ``investments of capital and administrative organisation'' can constitute authorship \cite{bently1994copyright}.

In January 2023, several academic publishers, such as the Science family of journals, updated their editorial policies to ban the naming of ChatGPT as author on papers, some going as far as to prohibit the use of AI technologies to generate content for papers altogether (though it is unclear how this can be enforced) \cite{doi:10.1126/science.adg7879}. Ostensibly, the practice of naming ChatGPT as an author stems from a desire to attribute its ``participation'' in the creative process. Yet this was not a safe or acceptable solution. A reason cited by publishers for this ban is that authors are accountable for the content in their papers, and ChatGPT is not an entity that can be held accountable to anything by anyone. Publishers also interpreted the use of ChatGPT output in a paper, as the (human) co-authors plagiarising text from ChatGPT, even it when named as an author \cite{doi:10.1126/science.adg7879}. Again, we witness a community response to the challenge of selecting one imperfect notion of authorship (and therefore creativity, or plagiarism) among a number of imperfect notions. The inclusion of ChatGPT as an author in this scenario was culturally unacceptable because of one particular interpretation of the author function.




A final example of how discourse, intent, and context determine the creative content of an object comes from our very own discipline of interaction design. In research-through-design, the aim is to produce knowledge by designing artefacts. The objects in isolation carry no knowledge content; the objects are simply design, not research-through-design. In research-through-design, the creative knowledge content of an artefact (which is judged by the community standards of peer review) depends on the rigour of the methods applied, and the associated discourse which elaborates the knowledge content and situates it with respect to related work \cite{gaver2012should,zimmerman2007research}.



With the view that creativity comes from authorial intent and discourse, when considering whether a particular episode of AI-assisted production is creative or not, we must not look only at the output itself. The output may be even be ``found'' or ``readymade'' from the training data, without attribution. A discursive assessment of creativity will consider, in what context is this output being used or presented? What is the authorial intent? What discourse is it facilitating? Who or what are we considering to be the author entity?

\subsection{Interpretation as creativity}
\label{sec:interpretation}

The previous discussion of meaning as constructed depending on context raises the question of who is doing the construction. This leads us to another family of ideas relating to the role of the observer (viewer, reader, consumer) of an object in determining its creative content. McLuhan puts it bluntly \cite{mcluhan1994understanding}: \emph{``A work of art has no existence or function apart from its effects on human observers.''}

The emphasis on the reader or viewer arose in part as a reaction to the status quo of literary and art criticism in the mid-20\textsuperscript{th} century, which focused so heavily on authorial intent and context, bordering on deification of the author. Barthes in his seminal \emph{Death of the Author} highlights the shortcomings of this view and offers a solution \cite{barthes2016death}: \emph{``it is language which speaks, not the author [...] to give a text an Author is to impose a limit [...] In the multiplicity of writing, everything is to be disentangled [...] writing ceaselessly posits meaning [...] but there is one place where this multiplicity is focused and that place is the reader''}.




Abstract expressionism aimed to elevate the process of production over the outcome. This also highlighted the artist's embodied presence in the work: the splashes of paint on a canvas by Pollock echo the swing of his arms. This elevated the artist above the viewer, which was an undesirable outcome for the counter-movement known as Minimalism. Artists such as Frank Stella and Kenneth Noland (and cognates in music: Steve Reich, Philip Glass, Brian Eno, etc.) sought to distance and even remove the hand of the author, using repetitive patterns, and a minimality of means, meaning, and structure, thus hoping to increase the involvement of the recipient in the work of art \cite{vaneenoo2011minimalism}. According to Bernard \cite{bernard1993minimalist}: \emph{``minimal art is aimed in part at achieving an impersonal quality, avoiding the depiction of personality that most minimalist artists felt had become entirely too explicit [...] to remove the accompanying interposition of artist's personality between artwork and viewer''}.




In literature the counter-movement against authorial intent became known as ``reader-response criticism'' \cite{iser1984reader}. Among the most vehement proponents were the American New Critics, who rejected the notion of authorial intent outright. The entire meaning of a text was in the text itself, and information from outside the text including and especially the author's intent, was irrelevant in interpretation. Having finished writing a text, the author is equivalent to any other reader, and has no special claim to understanding the work due to being its author. To value the author's intent above the interpretation of other readers was to commit the \emph{intentional fallacy} \cite{wimsatt1946intentional}, a literary foreshadowing of Dennett's intentional stance. The position of the New Critics is perhaps too extreme to defend, and counterexamples, such as the excerpt from Angelou in the previous section, clearly exhibit the need for extrinsic context for interpretation \cite{nuttall1968did}. The trap of assuming that the substantive content of a work can be derived purely from its form is the very same one we fall into when evaluating the creativity of AI based on its output. The point of reader-response criticism, however, is that even when extrinsic information is discounted, the same form can reveal different contents, different responses, depending on the reader's individual interpretation. The view of creativity as interpretation tells us that in order to evaluate whether something produced by AI is creative or not, we must ask: according to which interpretation of the output?

Yet elevating the reader and the importance of interpretation also has its own critics. According to critics of reader-response, allowing for ``interpretation'' to create content from form is a license to transform any work into anything else. At the extreme, interpretation renders meaningless the formal differences between texts; any text can be interpreted as having the meaning of any other text. Thus Sontag protests \cite{sontag1994against}: \emph{``Interpretation [...] presupposes a discrepancy between the clear meaning of the text and the demands of (later) readers [...] The interpreter, without actually erasing or rewriting the text, is altering it [...] interpretation excavates, and as it excavates, destroys; it digs `behind' the text, to find a sub-text which is the true one. [...] interpretation is the revenge of the intellect upon art''}. While expressed so poetically, the universalism of this position is as difficult to defend as that of the New Critics; Sontag appears not to acknowledge the role her own formidable and vengeful intellect plays in creating the ``clear meaning of the text''.

Many of these contradictory viewpoints are visible in the author-reader relations of J. K. Rowling, author of the Harry Potter series of fiction novels. The Harry Potter community looked to Rowling's authorial intent as an interpretive resource. When Rowling ``revealed'' that Albus Dumbledore, a main character, was gay (despite there being no evidence of this in the text), the community was eager to re-interpret the books with this lens \cite{irwin2015authorial}. This declaration of authorial intent signalled Rowling as an ally of the LGBTQ+ community. However, when Rowling's play \emph{The Cursed Child} began performances, the community accused Rowling of ``queerbaiting'' \cite{masad_2016}, which is to use the suggestion of homoromantic interest to appeal to the LGBTQ+ audience, while being non-committal enough to avoid being alienating to heteronormative society. For many in the LGBTQ+ community, the value of Rowling's authorial intent has been further undermined by her criticism of trans individuals, and her claims that trans-positive discourse is harmful to children, that trans activism is misogynist, and that trans women are a threat to ``biological'' women \cite{duggan2022transformative}. In response, queer communities are reclaiming the texts from Rowling's authorial intent, and re-queering them through fan fiction and community engagement.

The tumultuous relationship between Rowling and her community of readers shows that while literary critics are free to spill as much ink as they like for or against the intentional fallacy, for or against interpretation, the most forceful and consequential arbiter of the content of a work is the community that consumes it. Examples such as Rowling's echo in miniature the Protestant Reformation, perhaps the greatest of all community movements to reclaim interpretation from author-ity. The force of the community will also significantly shape the reception of artwork, literature, and any knowledge artefact produced with AI.

\subsection{Reuse as creativity}
\label{sec:reuse}

Recall the common position taken when assessing AI creativity: plagiarism is reuse without attribution. Reuse is permitted in creative works, but only with attribution: the taint of plagiarism negates any claim to creativity. Creativity and plagiarism are mutually exclusive. Thus one can only be creative either by avoiding reuse altogether, or by attributing every source on which one draws, which makes clear the \emph{novel} creative component of a particular work. But are these tenets universally applicable?

This section describes an alternative conception of creativity which holds that there is no such thing as creativity without reuse, and that exhaustive attribution is literally impossible. Moreover, norms around what constitutes ``proper'' attribution are not universal. In several historical circumstances, proper attribution has been deliberately avoided; there are practical and ethical reasons for ``improper'' attribution.

In literature the reuse inherent in every work is articulated by Barthes \cite{barthes2016death}: \emph{``a text is [...] a multi-dimensional space in which a variety of writings, none of them original, blend and clash. The text is a tissue of quotations drawn from the innumerable centres of culture. [...] the writer can only imitate a gesture that is always anterior, never original. His only power is to mix writings, to counter the ones with the others, in such a way as never to rest on any one of them.''}

Despite all text being a ``tissue of quotations'', we can still recognise new works as creative and novel. In \emph{The Anxiety of Influence}, Bloom examines the early 19\textsuperscript{th} century Romantics to develop a theory for how novelty can emerge from reuse \cite{bloom1997anxiety}. Bloom posits that all poets suffer from the anxiety of being derivative of precursor poets. However, Bloom observes that ``strong'' poets can respond in one of six ways to precursor poets, to create an original poetic vision despite the anxiety of influence. For example, a poet can offer a completion, or antithesis, to an idea ``started'' by a precursor poet.

It is important to note that the kind of reuse Barthes and Bloom talk about is fundamentally unattributable. It is a probabilistic, subconscious mixing of ``innumerable centres of culture'', of all sense-impressions and memories. If we take the Barthes/Bloom view of all work as fundamentally reuse, all authorship as fundamentally under influence, then pointing out the reuse nature of AI is redundant. The important question is how the AI output has responded to the training data it reuses, and whether it can be said to have constructed an original vision, e.g. by Bloom's criteria or by some other.

Not only do writers inexorably reuse ideas from other writers, but they also rely on the ideas of readers to help construct meaning. This is different from the earlier point about the role of reader interpretation. The theory of ``cognitive capitalism'' \cite{moulier2011cognitive} holds that any work depends on the ``general intellect'', a term mentioned in passing by Marx, which cognitive capitalists have expanded to mean the vast sets of knowledge and experience that are shared by members of a culture, and which artists draw upon. A comedian telling a ``knock-knock'' joke about waiting in line at a supermarket relies on the general intellect of a culture that is familiar with the form of such a joke, and also with supermarkets. Similarly, any text, artwork, or knowledge artefact draws on a basis of innumerable cultural, linguistic, and semantic connections -- most of which is not, cannot be, and need not be, attributed. Barthes extends this reliance on general intellect to the very words of the author's vocabulary \cite{barthes2016death}: \emph{``[Should the author] wish to express himself, he ought at least to know that the inner `thing' he thinks to `translate' is itself only a ready-formed dictionary, its words only explainable through other words, and so on indefinitely''}. In text, words are re-used resources, given meaning only by the network of associations in the general intellect of the reader (some semioticians may disagree, but a discussion of this is out of scope).

Words are fundamentally unattributable, as are concepts such as the knock-knock joke or the supermarket, nor do we wish to attribute them. If every work stands on a towering edifice of reused associations, we select only a miniscule fraction of these as worthy of attribution, depending on the context; this cultural delimitation of attribution reflects societal judgements of where ``value'' comes from in a work. These societal judgements are not moral universals, but are continually negotiated and renegotiated by different actors in the ``value chain'' of a knowledge work, to further private interests.

Numerous art forms are based on reuse. Collage, decoupage, and montage have been employed since the high middle ages. The art form of erasures, as in Tom Phillips' Humument \cite{phillips1987humument}, creates new texts by selectively erasing old ones. Reuse, when apparent, invites an automatic comparison between the original work and its reused form: this effect is exploited by subvertising, d\'etournement, and ``culture jamming'' to make artistic or political statements by corrupting and re-imagining public advertising and signage \cite{debord1956methods}. The attributional culture around each of these art forms does not require the authors of the source scraps of material to be attributed. No one demands attribution for the authors and photographers whose work appears in the magazine and newspaper clippings which constitute the Dadaist collage of Hannah H\"och, or the pop art photomontage of Richard Hamilton. These exemplify a distinct contextual, community-driven interpretation of creativity, plagiarism, and fair attributive norms.

Jazz improvisation includes the practice of ``quotation'': the inclusion of small, well-known melodic/harmonic motifs by other composers \cite{murphy1990jazz}. It is an encouraged and celebrated aspect of the art form, and functions as a way for the performer to demonstrate their knowledge of jazz repertoire, for them to converse and engage with other composers, and as a sign of respect and attribution to the quoted composers. Similarly, sampling practices in hip-hop \cite{schloss2014making}, and the practices of lyrical reference in rap music serve to create a dialogue between artists and also to acknowledge mutual creative influences. Unlike the world of textual attributions, which operates through names and bibliometric document identifiers, the attributional natures of quotation, sampling, and lyrical reference rely on the listener's knowledge and attitude. When a knowledgeable listener hears a quotation or a sample, it is heard not as an unattributed plagiarist reuse, but as a self-contained unit of conversation, reuse, and attribution. It is simultaneously a hat-tip to the prior composers and a wink at the audience. As the rapper Pusha T puts it so succinctly: \emph{``If you know, you know''}. This merging of citation and use, of quoting and attribution, was intolerable to Foucault, who nonetheless observed, and attempted to justify, that such merging was inevitable even in the highly formal attributional norms of scientific writing \cite{hess2006foucault}. Jazz quoting, hip-hop sampling, and rap references exemplify yet another distinct contextual, community-driven interpretation of how reuse relates to creativity, plagiarism, and fair attributive norms. Unfortunately, as we will see later, these norms come into violent conflict with the blunt instruments of copyright law.

Brueghel the Younger, a skilled painter but widely considered to be relatively unoriginal in comparison to his father, made a career out of copying and replicating his father's designs \cite{currie2012pieter}. Yet he was not a plagiarist: he did not present his own works as those of his father, he often subtly adapted the design to suit his customer's tastes, and he was extremely commercially successful and well-respected in his day \cite{kopenkina_2022}. Modern critics who have labelled him a knockoff or copycat are therefore inappropriately applying modern ideals around creativity and originality, modern notions of the artistic value chain alien to the ateliers of 16\textsuperscript{th} century Flemish painters, as though they are spatiotemporal universals.






In several historical circumstances, attribution has been deliberately avoided. Victorian women often published under masculine pen-names (e.g., Mary Anne Evans took the pen name George Eliot; the Bront\"e sisters and Louisa May Alcott also took masculine names) to increase the commercial acceptability of their work, and to avoid negative societal stereotypes attached to women writers. J.K. Rowling published a detective series under the pseudonym Robert Galbraith, to spare the series from the inevitable and unfair comparison to Harry Potter. Other writers published controversial or contentious ideas under pen names to avoid persecution or censorship (e.g., Voltaire -- already a pen name -- was known to deny the authorship of his controversial writings, preferring to attribute them to imaginary or sometimes real people). Eric Blair took the name George Orwell when publishing his memoir of growing up in poverty, to protect his family from embarrassment. Some men take on women's names when writing for an audience comprised of primarily women, in some cases multiple authors collaborate under a single pen name \cite{smith2006inkwell}. The point is that there are many legitimate reasons for subverting ``proper'' attribution which have no bearing on deemed creativity.

Another case where the assumed relationship between creativity, attribution, and plagiarism is challenged is in the forgery. The 18\textsuperscript{th} century poet Thomas Chatterton forged numerous poems in a medieval style, attributed to a 15\textsuperscript{th} century priest, which significantly influenced English, French, and German literature. Similar episodes of forgery include James MacPherson's ``Ossian'' and Edward Williams' ``Iolo Morganwg'' \cite{constantine2019possibilists}. There are many examples in the world of visual art, as well. Initially, these forgeries are considered creative on account of their content and their presumed authorship. Later, when the forgery is exposed, the creativity associated with these works does not disappear but is transformed. Today we celebrate the ingenuity of Chatterton and Williams, and appreciate the creative effort and talent required to masterfully reproduce the form of another artist, with the same conflicted fascination that we watch the protagonist in Frank Abagnale Jr.'s \emph{Catch Me If You Can}. It is not without reason that they are called con-\emph{artists}.


If we take the view that creativity is fundamentally reuse, we must recognise that every sentence we write, artwork we create, etc. is not only generated by the superposition of influences from innumerable prior works, but is also given meaning in the world only by reference to other works. We live in an era of unprecedented media availability and saturation, which McLuhan dubbed the ``electric implosion'' \cite{mcluhan1994understanding}. Filmmaker Kirby Ferguson's documentary series \emph{Everything is a Remix} recovers Barthes and Bloom in the digital age \cite{ferguson_2022}. It is easier than ever to find influences in our media environment, and more difficult than ever to avoid them. The anxiety induced in the Romantics by the passive, poetic influencer pales in comparison to the anxiety induced in modern society by the active, propagandistic social media influencer. \emph{``You Are Not a Parrot''}, declares \emph{New York} magazine, confidently summarising the view of linguist Emily Bender \cite{weil_2023}, who strongly resists any analogies between human language production and the ``stochastic parrot'' nature of large language models. Perhaps this view is defensible from a linguistic perspective, but it has been repeatedly repudiated from a literary perspective, a discipline which is decidedly closer to the questions of creativity and plagiarism.

It is impossible to determine every influence on our work and attribute them, so we settle on a culturally constructed compromise on the limits to what must be attributed. The result is that the overwhelming majority of influences that feed and enable any creative work remain unattributed. Thus the re-use and non-attributive natures of AI are not universal grounds to deny creativity, or accuse it of plagiarism. We must first ask: are we applying the standards of attribution appropriate to the context? What types of reuse are we (in)visibilising? Do the authors of sources desire attribution?

\subsection{Randomness as creativity}
\label{sec:randomness}

Randomness, or stochasticity, is an essential element of both the training and generation phases of AI. Critics who argue against the attribution of creativity to AI nonetheless note that randomness can be effectively used as a resource for producing creative work, as in Cage's ``aleatoric compositions'' \cite{leong2006randomness} and Eno's ``oblique strategies'' \cite{eno1975oblique}. In these practices, creativity comes not from the random process itself, but from the way in which the artist uses it to produce the resultant work. Blackwell writes \cite{blackwell2022coding}: \emph{``Tarot readings, gambling, and other entertaining performances, just like the compositions of John Cage, use random information as a starting point for human creativity. But [...] random information is not communicating anything [...] We can enjoy the performance of a Tarot reader, but the idea that random events have meaning in themselves is nonsense. [...] the same is true of the random processes that cause us to attribute `creativity' to AI systems.''}

Tarot, gambling, aleatoric composition, and oblique strategies are all practices which draw explicitly on randomness as a resource for creativity. Yet there are also many practices which, while they can be viewed as random, are interpreted by practitioners as channelling a different creative entity: the subconscious mind, the spirits of ancestors, or religious beings. 

The surrealist practice of ``automatic writing'', drawing upon the theories of Freud, aimed to produce writing without the intervention of the conscious mind \cite{muhl1930automatic}. Practitioners of automatic writing included W. B. Yeats and Arthur Conan Doyle. Automatic writers either held a traditional writing instrument such as a pen, and attempted to suppress their conscious processes while writing, or used a specially prepared board such as a planchette. Later interpretations of automatic writing aimed to channel not just the writer's subconscious, but to allow external spirits to guide the writer. This spiritual interpretation borrowed from practices in other cultures, such as Chinese Fuji spirit writing \cite{groot1964religious} for channelling various deities (first recorded in the 5\textsuperscript{th} century), medieval Christian glossolalia or ``speaking in tongues'' for channelling the holy spirit \cite{goodman2008speaking}, and American spiritualist talking boards (which evolved into the well-known parlour game of Ouija boards) for channelling the spirits of ancestors \cite{mcrobbie_2013}. In each of these cases, the exercise is not viewed as a human performance seeded by the random process, the random process is \emph{itself} the performance, given agency through concepts such as the subconscious or spirits.



There might be a tendency for readers operating within contemporary Western ways of knowing to respond to examples such as spirit writing and Ouija boards by denying their cultural interpretation of being ``extrinsic'' to the human actors: ``sure, spirit writers \emph{claim} they are not creating a performance, but in \emph{reality}, humans are still the true source of creativity, not any spirits.'' This is orientalist hubris. If we accept that definitions of creativity are ultimately contextual and community-driven, we must also be open to accepting community conceptions of the source(s) of creativity. The stochastic nature of AI is not universal grounds to deny its potential to be viewed by a practitioner community as a source of creativity.

\section{The form-content distinction}
\label{sec:form_content}

In previous sections, we have implicitly encountered the idea that there is a difference between form and content, or form and meaning. When we considered how authorial intent and interpretation can both affect meaning, it seems clear that the physical configuration of words on a page, or pixels on a screen, is only part of what determines the creative contribution (or plagiaristic nature) of a work. Here there are more community-produced norms regarding to what extent ``form'' determines content, and more fundamentally, what the ``unique form'' of a work even is.

\subsection{What is form?}
What do we consider to be the ``unique form'' of a text? Is it a certain configuration of letters or bits? Is each different occurrence of the text, whether on screen as an array of coloured pixels, or in ink printed on paper, a different form? In general we would not consider two printed copies of the same text to be different forms, despite the innumerable minute physical differences between the copies, such as being printed on different paper, with slightly different ink, etc. These differences are not relevant to our typical determination of textual form, which generally relies on an abstract notion of an arrangement of characters. A notion of ``unique form'' centres around relevant and irrelevant differences. Unique forms are separated from each other only by relevant differences. Forms separated only by irrelevant differences have the same unique form.

But there is no notion of form universal to all texts. Change the line breaks in a news article and it remains the same. Do the same to a poem and the meaning changes. Change the typeface of a school homework essay and it remains the same. Do the same to a magazine advertisement and it is transformed. Some genres of literature take this formalist emphasis to the extreme: ``Concrete poetry'' celebrates the typographic form, showing how the arrangement of letterforms can be used as a visual material, quite distinct from the language semantics that arise when arranged into known words. As Draper puts it \cite{draper1971concrete}, \emph{``the shaping of patterns takes preference over communication''}. Similarly, ``sound poetry'' focuses on the architectural configuration of phonetic speech sounds, rather than the formation of words \cite{perloff2009sound}. Concrete and sound poetry can be viewed as prioritising ``form over content'', yet they can also be viewed as showing how a different kind of content can be created out of form. This fascination for form-content subversion is not limited to obscure corners of early 20\textsuperscript{th} century poetry; to show its enduring popularity one need only note the viral success of the Netflix series \emph{Is It Cake?}

In philosophy the issue of ``type-token'' distinction is related \cite{sep-types-tokens}. The problem is that literary and art critics (or critics of AI output) often talk about a work as an abstract form, or a type. Yet ``types'' do not physically exist, only tokens -- manifestations that we can perceive. Philosophical scholars have variously characterised types as sets, kinds, or laws, and have argued for both their existence and non-existence. The question for AI is this: each output ``token'' (in the philosophical sense, not the natural language processing sense), i.e., each time an output is displayed, is generally not considered to be a novel creative act. If an AI's output text is copied and pasted form one application to another, or printed, these various instances are generally all considered to be tokens of the same original ``type'', an attitude we inherit naturally from 21\textsuperscript{st} century knowledge work norms. In doing so we are implicitly choosing some notion of ``form'' which may not be universally shared by all human interpreters or appropriate for all contexts. This is relevant when the work is being scrutinised for creativity or plagiarism. If we choose a notion of form that is sensitive to line breaks, to typeface, to materiality, etc., then these must be carried over from the original token into any subsequently distributed copies, and we must be aware of how the processes of copying and distribution might introduce relevant formal (potentially creative) differences.








\subsection{How much does form determine content?}
Once a community has agreed suitable norms around what constitutes form, there is the question of to what extent this form determines content. In art and literary criticism (except with procedural and conceptual art), despite the enormous influence of authorial intent and interpretation, form is generally considered paramount to content. Change a line of a poem by Wordsworth, or a soliloquy by Shakespeare, and the entire content is transformed. Retelling the story of Romeo and Juliet in my own words is to not convey Shakespeare at all, and would not be considered plagiarism in the strictest sense. A rose by another name is another thing. A restatement is a reinvention.

Contrast that to the way in which form is understood to relate to content in academic writing. In this paper, like in all academic writing, an idea is presented that is understood to be quite distinct from the words used to convey it. This allows the community to defend against a different kind of plagiarism during peer review: \emph{``This idea was already published by X et al''}. A restatement is not a reinvention.

Art and literature harbour both extremes of form-content determination. On one extreme, where content is an idea quite abstracted from form, is popular media. The popular conception of originality tends to skew in preference to evaluating an abstract notion of content over a precise form. Critical television viewers, readers, and moviegoers tend to reject storylines and tropes that are borrowed wholesale from other popular works with which they are familiar, recognising that the fundamental ``idea'' underneath is what counts, not the way in which it is expressed. On the other extreme, where content is entirely determined by a precise form, are writers like Walter Benjamin. In \emph{The Translator's Task} \cite{benjamin2021translator}, Benjamin explains that a translation -- whose entire aim is ostensibly to achieve a shift in form without a shift in content -- is fundamentally impossible: \emph{``translation is merely a preliminary way of coming to terms with the foreignness of languages to each other. A dissolution of this foreignness [...] remains out of human reach [...] the relation between content and language in the original is entirely different from that in the translation [... which] remains inappropriate, violent, and alien with respect to its content [...] from outside it, facing it, and without entering it, the translation calls to the original within, at that one point where the echo in its own language can produce a reverberation of the foreign language's work''}.

The line between form and content is precisely the one skirted by knockoff brands, which evade legal action through clever modifications of the legally formalised elements of brand identity, which differ sufficiently from the original to render them nonequivalent according to formal legal comparisons. On the other hand, the similarities they retain allow them to still benefit from consumers' mental association with the original brand.


Political activism has long exploited the ability to separate content from politically-controlled form. In 2022, Chinese lockdown protesters were warned by the police not to ask for an end to lockdowns. The crowd responded \cite{wang_2022} by sarcastically chanting \emph{``Continue lockdowns! [...] I want to do Covid tests!''} Sarcasm reverses the assumed content of form. Other protestors held up blank signs in reference to a Soviet-era joke \cite{adams2005tiny, che_chien_2022}; \emph{``They know what they want to express, and authorities know too, so people don't need to say anything. If you hold a blank sheet, then everyone knows what you mean''}. Context has the power to synthesise content from an entirely absent form.

The interplay of form, context, intent, and interpretation in producing content are key not just to political communication, but innumerable instances of ordinary, everyday communication (as noted in Grice's theory of ``implicatures'' \cite{grice1975logic} and Miller's theory of explanations \cite{miller2019explanation}). Indeed the study of this interplay is an entire sub-field of linguistics \cite{yule1996pragmatics}. 

In AI, influence of an idea in the training data may reveal itself as what appears to be the same idea in content, but not in form. Would this be considered plagiarism or creativity? A restatement or a reinvention? The answer depends on which perspectives on the form-content distinction we apply when evaluating the output. Are we reading it as a poem, or a scientific paper? Are we applying a popular conception of originality, or Walter Benjamin's?

\subsection{Labour, mechanisation, and creativity}

Culturally produced form-content distinctions are nuanced and resist generalisation, but I would cautiously suggest a recurrent pattern: form-content distinctions (and therefore, distinctions of creativity and originality) fall along the lines of mechanical reproduction. Where we see the labour of the human hand, we see a difference in content. If a work is copied artfully by hand, the human labour involved often grants it a new ``aura'' \cite{benjamin1935work}, an originality of its own. When the medieval scribes copies a manuscript, or the sculptor's apprentice copies an original marble by his master, the copies have intrinsic creative value despite their ``formal'' equivalence to the original. But when a text is copied by print, or a sculpture produced through a 3D printer, we see it as the same work as the original. Today we are much less likely to consider a hand-copied text to be different from the original than our medieval ancestors, because our cultural attitudes to the value chain in text production have been conditioned by centuries of the mechanisation of print.

The general (though not universal) principle is this: the lower the marginal human labour of reproduction, the less we distinguish the new forms as unique types, the more we see them as tokens of the same general type. Mechanisation turns formerly relevant differences in form, differences which arise or can be manipulated through the mechanical process, into irrelevant differences. I believe the same principle is at play in the Japanese manga example at the start of the paper: ``classical'' fan art is culturally acceptable because copying involves labour. AI-generated fan art is not, because the form of copying it enables is not laborious in the same way. Goldsmith notes that the huge quantitative increase in the speed of copying enabled by computers resulted in the qualitative departure of conceptual writing from earlier forms \cite{acker2011against}: \emph{``previous forms of borrowing in literature -- collage or pastiche, taking a sentence from here, a sentence from there -- were predicated on the sheer amount of manual labor involved: to retype an entire book is one thing, and to cut and paste an entire book is another. The ease of appropriation has raised the bar to a new level''}. 

Similarly, the launch of DeviantArt's DreamUp AI art generation service upset many artists due to its opt-in default \cite{edwards_2022}. Commentary around the community reaction rejected the comparison of AI art generation to collage, feeling that copying the style of an artist was qualitatively different from collage. Artists' own conceptions of the fundamental injustice again centred on the labour requirements of creating art in this new way: it feels more unjust, more like copying, when it is made easy in comparison to familiar manual methods. Artist Kelly McKernan said \cite{woods_ma_2023}: \emph{``I almost barely make rent most months. I will spend, you know, 30 hours on a painting, and I won't see any money from that until it sells. [...] There's more and more [AI-generated] images that I can see my hand in, but it's not my work. I'm kind of feeling violated here. I'm really uncomfortable with this''}. Technological advances have repeatedly thrown into crisis the ideals of creativity and labour in various artistic communities, forcing their re-evaluation. In section~\ref{sec:AI_natures} we will inspect how creativity is being re-evaluated in response to the challenge of AI, namely, as a shift from material production to critical integration. 



\section{The challenges of intellectual ``property''}
\label{sec:intellectual_property}

It might seem helpful to look at legal definitions of originality and creativity, e.g., those referenced in copyright law, to help understand whether AI can be considered creative. The objective of this section is to explain why it is \emph{not} as helpful as it may seem. The relatively short history of intellectual property (IP) rights is essentially that of the privatisation of thought as a capitalist instrument of control. IP law has simultaneously been successful in furthering private interests, while failing to capture a logically consistent notion of creativity. 



IP consists of legal devices such as copyrights, trademarks, patents, etc. These are a relatively recent invention. For example, the first recognisably modern copyright law was the English Statute of Anne of 1710, which granted publishers the exclusive right to print a work for 14 years after its first publication. It was introduced not out of a desire to empower authors, as it was cleverly pitched at the time, but rather to further the interests of powerful centralised printing companies against cheaply made provincial and foreign books. Patterson observes \cite{patterson1965statute}: \emph{``The Statute of Anne [...] developed by and for publishers, was clearly a publisher's right, not an author's right''}. The rights granted implicitly to authors in this statute were not recognised in English common law until 1774, sixty-four years later. Similarly, we can expect a new generation of legal devices invented to manage the IP rights regarding AI output, and these will not at first protect the entities we subsequently come to understand as authors, but rather protect the most powerful lobbying voices.

Just as the privatisation of land forces people to sell labour in order to eat \cite{sedgewick2021coffeeland}, so the privatisation of thought functions as an incentive to work in a knowledge economy. Pasquinelli explains the motivation to pursue stronger intellectual property regimes bluntly \cite{pasquinelli2009google}: \emph{``copyright is one of the strategic evolutions of rent to expropriate the cultural commons and reintroduce artificial scarcity. Speculation then is directed toward intellectual property, forcing artificial costs on cognitive goods that can paradoxically be reproduced or copied virtually for free.''}







\subsection{Intellectual property's dual agenda and reliance on notation}
IP rights are a ``negative right'': broadly speaking, the right to prevent copying of creative ideas. This is a Western idea prioritising the individual \cite{smith2021decolonizing} based on the theory that if you protect people's ideas, they will innovate more, because they stand to protect their financial gains. On the other hand, progress is cumulative and ideas build on previous ideas, so one might expect that society will innovate more if there are fewer IP laws. IP law therefore aims to strike a balance between these two opposing tendencies. Economic models and empirical analyses generally come to the intuitive conclusion that some IP rights can improve innovation, but others can hinder it \cite{neves2021link, gangopadhyay2012does}. IP rights therefore function more as a governmental lever or dial for regulating innovation, like the interest rate of a central bank, rather than as a quest for the legal enshrinement of principled notions of creativity and ownership. Except that in order to do so, it has entangled itself in this same quest.





The idea that one owns what one produces through one's own labour goes back to Locke \cite{day1966locke}: \emph{``Whatsoever, then, he removes out of the state that nature hath provided and left it in, he hath mixed his labour with [...] and thereby makes it his property''}. The related labour theory of value developed by Smith, Ricardo (and then Marx) appears somewhat later \cite{day1966locke}. This is complicated by intellectual property in two ways: first, the ``labour'' involved in producing an idea is undefinable and unquantifiable, unlike the labour of producing material goods. Secondly, labour alone appears insufficient to qualify for ownership: if I labour to copy a poem written by someone else, cultural norms around form and content in poetry dictate that the idea for the poem does not now belong to me. Thus copyright law in general rejects this ``sweat of the brow'' doctrine and instead has to define a notion of originality, i.e., creativity \cite{ginsburg1992no}. This dual agenda of regulating innovation and defining creativity is the source of many contradictions.

The main source of problems in the agenda of definition is its reliance on \emph{notation}. For an idea to be apprehended by the legal apparatus, for any judgement to be made regarding the equivalence or nonequivalence of ideas, they must be notated in a way that facilitates comparison. Selecting and developing a suitable notation for the legal codification of ideas is the same as selecting a notion of unique form, of deciding which differences are relevant and which are irrelevant. Any particular choice of notation will inevitably highlight certain aspects of ideas while ignoring others.


In the case of music copyright, courts note that a musical idea must be written down in staff notation \cite{latham2003newton}. This inherits from the specific ideals of the European Romantic composers of the 19\textsuperscript{th} century, that the composer's notation expresses the true ``intent'' of the music. Yet most genres of music across human cultures and history, including the vast majority of contemporary music, is not composed or amenable to notation in this way. This leads to absurdities such as prohibiting jurors from listening to the music in question \cite{davis20189th}, only looking at staff notation (which is transcribed by a court-appointed notator, not the artists themselves) for determining equivalence and nonequivalence of musical ideas. 

In response to the challenges of notation, courts will also employ ``ordinary observer'' or ``lay listener tests'', which amounts to playing two pieces of music to the jury to assess if the compositions ``feel substantially'' the same, while recognising the primacy of the notational tests of equivalence. This leads to baffling and contradictory procedures, such as the following \cite{davis20189th}: \emph{``When the case returns to trial, the jury will continue to not be able to hear [the] original album recording [...] in order to judge its substantial similarity [...] The jury will be able to hear the recording in order to weigh access, but there might be measures in place to make sure the jury doesn't listen too closely''}. Unsurprisingly, these tests draw criticism from musicians and legal scholars alike, who note that lay listeners apply inconsistent and easily biased criteria to judge musical similarity \cite{lemley2009our, palmer2015blurred, lund2011empirical}. 

Copyright law includes many exceptions and compromises to paper over questions of form and content, such as the separation of recording, composition, and performance rights, and the ``idea-expression'' dichotomy \cite{newman1999new,vezzoso2012copyright}, all of which are further problematised by new modes of reproduction and performance enabled by digitalisation. Copyright also includes various exceptions to cater for the contextual nature of reuse (such as the ``fair use'' doctrine which enables reuse for purposes such as education). These exceptions and contingencies are the result of intellectual property law attempting to simultaneously act as a governmental lever for innovation and as an arbiter of originality. The frail nature of copyright comparisons is also vulnerable to absurd attacks such as Riehl's ``All the Music'' project \cite{riehl_2020}, which has mechanically generated all possible melodies 10 notes long and released it into the public domain, to help musicians defend themselves against allegations of copying. 

The dual agenda of defining creativity and regulating innovation requires arbitrary decisions to be made which can be influenced by private interests. For example, time limits. Copyrights are not held indefinitely: in the USA for instance, copyrights expire 70 years after the death of the author. In what way has the creative nature of the work changed on the first day of the 70\textsuperscript{th} year? The number 70 is arbitrary, it has been changed several times, and is manipulated by commercial interests. There is some evidence that the extensions to US copyright protection periods in 1976 and 1998 were influenced by lobbying from Disney, whose copyright on Mickey Mouse was about to expire in each case \cite{moffat2004mutant}. Copyright draws arbitrary lines in space as well as time: they are unevenly translated between the jurisdictions of different countries, and they often have different notational practices for determining equivalence. An idea that is derivative on one side of the border magically becomes original on the other side.

Another absurd episode in copyright law is the  case of the infamous ``monkey selfie'' \cite{rosati2017monkey}. This was a dispute between British photographer David Slater and the Wikimedia Foundation and People for the Ethical Treatment of Animals (PETA). The dispute, which lasted for seven years, centered around whether Slater held the copyright to a photograph of an Indonesian crested macaque that the macaque had ``taken'' of itself while manipulating Slater's camera. In the final ruling Slater was granted the copyright not because the courts determined that he was the source of the creativity and originality in the image (it was clear that PETA was not actually interested in the question of creativity but rather sought to raise its own profile through a sensationalist lawsuit; we will see similar attempts to grab headlines from companies eager to defend the ``rights'' of AI), but on the rather less interesting technicality that non-human actors are not entitled to copyright protection. But in that case, corporations must not be entitled to copyright protection either, because they are not humans. Yet corporations clearly do hold copyrights. This apparent contradiction (and many others) is resolved due to the legal standing of corporations as ``artificial persons''. This artificial personhood is what allows corporations to be held accountable for actions, and some call for AI systems to be given similar personhood to improve its accountability (though there are criticisms of that approach \cite{bryson2017and}). Depending on whether a particular AI system has been given legal personhood, it may or may not be entitled to copyright protection under the current legal regime, regardless of its creative role.

In the USA a war over electronic book lending is being waged between librarians and publishers. With physical books, the library purchases the book and then owns it forever. But they can only lease electronic books for a certain period of time, after which the lease needs to be renewed. This is an exorbitant recurring expense which many libraries cannot afford. The publishers' argument is essentially that e-book lending makes it much easier for the public to borrow books, which reduces sales. Yet this contradicts, or at least undermines the spirit of the ``first sale'' doctrine of copyright, which allows owners of copyrighted works (e.g., libraries) to lend, sell, or share their copies without requiring permission or paying additional fees \cite{jenkins2014last}. The vastly reduced costs of replication and distribution throw into crisis these notions of copyright, exposing its basis as a practical, commercial mechanism, rather than any universal notion of creativity and ownership.


\subsection{Societal harms of legal definitions of creativity}
The use of copyright to extract value from the creativity of minorities and to deprive them of credit, compensation, and control, is well documented \cite{lester2013blurred}. Conversely, challenges to the notationally-privileged elite always emerge from marginalised communities. The practices of DJ-ing and turntabling, from which the referential practices of hip-hop and rap emerged, were forged in the devastating fires of the South Bronx in the 1970s \cite{katz2012groove}. Artists with no access to musical instruments or training made virtue of necessity and turned their playback machines into musical instruments. They echoed the emancipatory movements made by the cool, improvisational, quotational jazz of decades earlier, which drew on black literary and musical traditions \cite{murphy1990jazz}. When marginalised communities incorporate AI into their creative processes, we may see similarly disrespectful and appropriational attitudes from the privileged elite. The stochastic reuse nature of AI may be held against marginalised communities in the same way that sampling or jazz quotation struggles to defend its creativity in notationally-biased courtrooms. We may see transnational conflicts with xenophobic undertones, as exemplified in the attitudes of Western pharmaceutical giants seeking to suppress the rights of Indian pharmaceutical companies to manufacture affordable medicine \cite{grace2004effect}, or the unfair generalisation from a small set of bad actors to the branding of the entire Chinese technology industry as being built upon IP theft \cite{lee2018ai}.





Copyright law is further complicated by the fact that copyright of not just original but also ``derivative works'' is granted. This opens up avenues for rent seeking. Many copyright lawsuits are brought not by artists but by investment firms who have bought the rights and seek to extract rent by proving ``derivation'', essentially the same business model as patent trolling. This has deleterious effects on entire creative industries. The aftermath of the 2015 \emph{Blurred Lines} and \emph{Stay With Me} copyright lawsuits has profoundly affected music production \cite{sisario_2019}. Many artists report feeling nervous while composing, and fearful of trespassing into what they have now been conditioned to see as the territory of previous artists. \emph{``I shouldn't be thinking about legal precedent when I am trying to write a chorus''}, said songwriter Evan Bogart \cite{sisario_2019}. New pre-emptive citational practices have emerged, occasionally going as far as to name influences as co-artists. Beyonc\'e's 2022 single \emph{Break My Soul} credits Robin S.'s performance of \emph{Show Me Love}, despite clear and notationally-defensible differences between the two songs \cite{cantor_2022, harding_2022}. Additional credits continued to be added and removed over time \cite{juzwiak_2022}. Cautious over-attribution of this kind, and the consequential payment of fractional royalties, seems to be anticipating, and hoping to defuse, the crude approximate comparisons of the lay listener test. Foucault noted how intellectual property laws in a slightly different way, exerted a chilling effect on authorship, as when granted a text as his ``property'' by copyright law, an author needed also to bear the risks, including punishment for expressing transgressive views \cite{foucault2017author}: \emph{``[The text's] status as property is historically secondary to the penal code controlling its appropriation [... prior to copyright laws] books were assigned real authors [...] only when the author became subject to punishment [... but] when a system of ownership and strict copyright rules were established [... it restored] the danger of writing''}. The takeaway for AI is this: community-driven norms for attribution can shift in response to the threat of legal action. Attributional practices can be distorted by perceived threats and thus become detached from notions of creativity and originality.

As far as musical creativity goes, the effort to establish universal, fair, and logically consistent criteria for originality, equivalence, and nonequivalence have largely failed. As the creative content of AI generated works begins to undergo the same legal scrutiny, we can be assured that the same jumbled mess of notations and tests will continue to produce non-answers, and pressurise creative communities to second-guess their creative workflows and attributional practices.






\subsection{Plagiarism or market-making?}
As noted in the previous section on reuse: in the world of information, a curatorial voice (Bush's trail blazer) offers more value now than ever before. The same has long been true of physical goods. Shaviro \cite{shaviro_2008} notes that Walmart, the world's largest employer, \emph{``focuses entirely upon circulation and distribution, rather than upon old-style manufacturing — showing that the sphere of circulation now (in contrast to Marx's own time) plays a major role in the actual extraction of surplus value''}. A popular ``get rich quick'' scheme is drop shipping (e.g., \cite{hayes2013ultimate}). To start a drop shipping business, one sets up a digital storefront through which customers can order various goods. When an order comes in, the drop shipper purchases the item ``just in time'' from another store, and orders it to be shipped to the customer. This business works by price arbitrage: the drop shipper simply purchases the good at a lower price, but sells it at a higher price. The supplier is (usually) a Chinese producer with a listing on a Chinese-language online retail store such as AliExpress. The drop shipper's storefront is designed to be attractive and claim a certain level of luxury, often set up in English and aimed at European and North American markets. Unlike a retail distributor, the drop shipper's infrastructure and investments are minimal: there are no real estate costs, no costs for maintaining inventory due to the just in time approach, and negligible labour and operating costs after deploying the online store. Entire businesses are built on the premise of labelling cheap mass market commodities as luxuries, such as the successful Daniel Wellington watch brand \cite{pulvirent_2015}. 

In (loose) economics terms, drop shippers can be interpreted as providing value because they are market-makers \cite{amihud1980dealership}. They are benefiting from the failure of the market to directly connect customers to suppliers and offering an interface. On the flip side, they can be viewed as exploiting the ignorance of the consumer, who could, with a little effort, order the product at the lower price themselves, and the ignorance of the supplier, who could, with a little effort, set up their own similar storefront and charge the premium price themselves. In a similar manner, the copyright infringer, e.g., the seller of unauthorised prints of books, or bootleg CDs, or pirated movies, could be viewed as solving a market failure to bring a certain original work to a certain viewer, or could be viewed as exploiting the ignorance of the viewer and of the original source. In legal terms, they are completely different scenarios. Drop shipping is legal and seen as solving market problems. Copyright infringement is illegal and seen to be exploitative. In terms of helping transport information from source to end-consumer, they are identical. The difference arises in the \emph{consequences} of these activities for the various stakeholders involved. For instance, drop shipping deprives the source of only part of their potential revenue; piracy deprives them of all of it. The consequence for AI is this: large language models and image models can be viewed as a newly efficient method for connecting the end-user with a space of information defined (partly) using a training dataset. In some cases this facilitation of information flow could be viewed as helpful market-making. In others it could be viewed as exploitation.

\textbf{In summary}, this section has discussed the problems created by the dual agenda of intellectual property rights and its reliance on notation, and the numerous and unsatisfactory compromises necessary to make such an agenda work in practice. For AI, the implication is that relying on legal apparatus to help define creativity and plagiarism is subject to all the same compromises, and simply furthers the private interests of powerful actors. As designers of socio-technical systems, to appeal to legal arguments is to knowingly abrogate responsibility to a failed project.

\section{Shifting community norms for creativity in AI-assisted knowledge work}
\label{sec:AI_natures}

So far we have considered many different conceptualisations of creativity, of form and notation, and the challenges of legal definitions. I have suggested two patterns: first, that community is the ultimate arbiter of creativity, and second, that community conceptions of creativity often fall along the lines of mechanical (re)production. A rapid fall in the cost of mechanical production in some parts of the creative process causes communities to reconceptualise activities such as art and writing to identify new sources of human value and new loci of creativity. This section will briefly outline some emergent patterns for where these new loci might sit according to practitioner communities, based on studies of AI use in art and writing.

\subsection{From production to critical integration}
\label{sec:critical_integration}

AI models mechanise and thereby vastly reduce the cost of the production of information, such as text or images. With further advances in AI, the marginal human labour of material production, of physically writing a text, or creating an image, will approach zero. When it is mechanised, the process of material production will cease to be viewed as creative. Instead, what we can observe is a shift in knowledge work from production to \emph{critical integration}. The output of AI systems will need to be \emph{integrated} into a wider workflow involving human action. Creative labour is therefore expended in deciding where in the workflow to use the productive power of AI, how to program it correctly (e.g., in the current generation of large language models this is done through ``prompt engineering'' \cite{liu2022design}), and how to process its output in order to incorporate it. This integration must be \emph{critical}, meaning that creative labour is expended in qualitative and expert human assessment of the output, e.g., generated text might be checked for factual accuracy, or generated code might be checked for correctness. 

Some have likened this workflow to a ``sandwich'' \cite{smith_2022}, where the human work of prompting and editing surrounds the AI generation process. Critical integration generalises the sandwich, if you like, to the entire ``double loop'' \cite{argyris1977double} of critically reacting to individual instances of AI use, and readjusting entire knowledge workflows in response to meta-observations and hypotheses about the roles AI might play. The concept of critical integration can be illustrated through three studies, in creative writing, in visual arts, and in programming.

First, Singh et al. studied how creative writing workflows shift after the introduction of AI assistance \cite{singh2022hide}. In their study, AI assistance was available continuously as the authors wrote and took multiple forms: two different types of textual suggestions, and suggested images and music retrieved from online databases which were displayed/played ambiently within the editor. For us, the most important observation they make in this study is that AI-assisted writing consisted of ``integrative leaps'', defined as \emph{``the different kinds of interpretation and expression involved in incorporating aspects of suggestions into the developing story''}. They identify multiple axes of integration such as indirect-direct (to what extent is an AI suggestion used as-is, without modification), and exploratory-confirmatory (to what extent is an AI being used to continue the current narrative, versus as a tool for exploring alternative narratives). Participants were willing to attribute creativity to the AI system (e.g., \emph{``It was surprising to see the intelligence of the AI and the creativeness of the suggestions''}), yet still felt ownership of the final text, because of the numerous authorial decisions required in the course of critical integration (e.g., \emph{``it helps me find an idea, but I was the one who developed the story and make the story coherent''}).

Second, Ploin et al. studied how AI tools affected the workflows of a cohort of visual artists \cite{Ploin2022}. They found that artists engaged in five new activities: studying AI to gain a better technical understanding, selecting/building/combining models, building datasets, training models, and curating outputs. Artists made critical and artistic judgements at every stage of the process. The analysis reveals a new community norm around creativity emerging, which permits the creativity of production to be attributed to the system. As one artist put it: \emph{``In the process, the model was entirely 
more creative than a human. It created images [...] I can’t create''}. Yet the role of critical integration, particularly integration of artwork into the human-societal discourse of art, is a newly strengthened creative responsibility of the artist. The artist continues: \emph{``The pictures of the petals are beautiful [...] but it’s only made more meaningful by [...] a 
human creator to contextualize and understand the present moment, because art is created for people in a specific cultural moment''}. Just as with the Singh et al. study, the AI in these workflows is viewed as its own intrinsic source of creativity while \emph{simultaneously} the artist performs the creative work of critical integration. As the authors note, \emph{``creativity is an easier target than art''}. Some creative work can be fairly attributed to AI, but more creative work is required to turn it into a knowledge product. Another artist said: \emph{``I believe that these algorithms are and can be creative. [...] But there is a big distinction between that and making art, or art that is interesting or valid. That requires a lot of intentionality''}.

Third, Sarkar et al. studied reports from programmers using AI assistance for writing software code \cite{sarkar2022programmingai}. They too observe how the creative work of writing code changes with AI assistance. It shifts away from directly determining what character sequence expresses the programmer's intent and then physically typing it out. It moves towards determining suitable prompts (\emph{``breaking down a prompt at the `correct' level of detail is also emerging as an important developer skill''}), identifying appropriate opportunities to use AI assistance in the workflow (\emph{``this requires users to form a mental model of when [AI] is likely to help them in their workflow. This mental model takes time and intentionality to build [... programmers are] constantly evaluating whether the current situation would benefit from [AI] assistance''}) and then critically evaluating and assimilating the output once generated (\emph{``developers need to learn new craft practices for debugging''}).

Besides the studies mentioned above, there have been many other investigations of how contemporary generative AI supports creative and knowledge work \cite{mirowski2022co, chung2022artist, miller2019artist, van2021ai, yang2022ai, dang2022beyond, gero2022sparks, suh2021ai, oppenlaender2022creativity, inie2023designing, caramiaux2022explorers}. Most of these studies provide empirical evidence or analytical support for varying degrees of shift towards critical integration and away from production, and show this shift to be a general pattern across many different types of creative and knowledge work.

\subsection{Implications of critical integration}
\label{sec:critical_integration_implications}

In the wake of the release of ChatGPT, many were concerned about the impact of AI on education, since writing textual answers and essays forms so much of the ``output'' required by students over the course of their learning careers \cite{rudolph2023chatgpt}. What is the point in learning to write if AI can do it for you? Some lamented this development, while others were unruffled, pointing to the effect of calculators on mental arithmetic. The shift to critical integration suggests that perhaps editing and proofreading an AI-generated essay becomes the new way of testing student skill, rather than the material production of the words of the essay.

In the interim, what might be the implications of the shift to critical integration for labour process issues, class relations, and the agenda for HCI research and practice? These are important questions which are largely out of scope for this paper; the focus of this paper has been to explain the nature of critical integration and to broaden the scope of discourse around AI and creativity in knowledge work. However, it is worth briefly touching upon the most salient issues. Braverman exposed the tendency of capitalism to destroy craftsmanship and degrade work through Taylorism \cite{braverman1998labor}, extending a Marxian analysis to the 20\textsuperscript{th} century. Greenbaum suggested that Computer-Supported Cooperative Work (CSCW) research and technologies had been complicit in this degradation, by enabling the transition from secure, well-paid ``brick and mortar'' jobs to low-paid and precarious jobs that were nevertheless highly skilled and challenging \cite{greenbaum1996back}. As an antidote, Greenbaum suggests that designers and researchers explicitly consider the consequences of a design agenda on labour, wages, and organisational structures, using labour studies as a guide. He further suggests that they ought to be held accountable for these consequences. 


Just as Greenbaum claims CSCW research has contributed to a Marxian breakdown of knowledge work and created new class struggles and consolidation of capitalist power, so too do we need to consider whether we are enabling these outcomes with generative AI, or at least incorporate the labour process perspective from the outset. Moreover, just because the community-produced definitions of creativity might be shifting due to critical integration, and just because material production can be automated, it does not follow that everyone within a community wants it to be automated. The implications for the class identity of knowledge work in response to AI is too broad to be treated properly in this paper \cite{karakilic2022humans,marks2009stuck}, though it is of great importance to explore in future work.


However, we are still in a transitional phase, and it is unclear how the technology will evolve and how society will evolve in response. There are numerous examples of workflows in the early industrial revolution that were not fully mechanised and were repeatedly reconfigured over the course of decades as it became possible to automate more and more of the workflow (or alter the manufacturing objective to require less human intervention) \cite{crawford_2018}. McLuhan makes this point about printing: \emph{``Typography was no more an addition to the scribal art than the motorcar was an addition to the horse. Printing had its `horseless carriage' phase of being misconceived and misapplied during its first decades, when it was not uncommon for the purchaser of a printed book to take it to a scribe to have it copied and illustrated.''}

It is harder still to anticipate how the fundamental enterprise of knowledge work will shift in response to the challenge of AI in the longer term. Perhaps critical integration is merely a stepping stone, and future iterations of technology will successively erode the need for human involvement in knowledge workflows in ever-increasing proportion. In which case conducting knowledge work ``by hand'' may for most become an activity for pleasure and recreation, or an issue of class activism.

\subsection{Resisting mechanised convergence through cultural reversal}
\label{sec:convergence}

In 1983 Bainbridge noted the ironies of a system that automates ``normal'' operations while relying on human operators to step in to handle exceptional situations and mitigate system errors \cite{bainbridge1983ironies}. This is the situation we find ourselves in today, a complete reversal of the 19\textsuperscript{th} century motivations for developing computing machines (articulated by Babbage and others \cite{snyder2011philosophical}), which was to overcome human errors. We have returned from ``humans make errors, so computers step in'' to ``computers make errors, so humans step in.'' This reversal was perhaps inevitable as we charged computers with an increasingly broader and more complex set of tasks. McLuhan notes many such examples of reversals that occur as a technology matures or intensifies \cite{mcluhan1994understanding}.

One tendency of mechanised production is the convergence of expression to known and fixed forms. The printing press helped to standardise spelling (though it is debated to what extent \cite{howard2006early,brengelman1980orthoepists}), much as spell check does today. Mass manufacturing standardised many previously bespoke commodities, such as clothing. But when this tendency for convergence is taken to an extreme, it becomes repulsive. A clear example is the outcry surrounding Huawei's ``Moon Mode'' feature, which purported to use AI to improve the clarity of pictures of the moon taken using the camera on the Huawei P30 Pro smartphone. However, it emerged that the feature worked by merging the user's photograph with previously captured higher resolution imagery \cite{brown_2019}. On paper, it seems like a good idea: it is \emph{the} moon, after all, and it is the same for everyone. But the feature received severe criticism for failing to note a basic sentiment that accompanies picture taking: the pride of personal possession. This is why millions of tourists take terrible, blurry photographs of the Taj Mahal and the Eiffel Tower each year despite the fact that high resolution imagery is freely available online. Being connected to the act of taking the photograph is more important than the photograph itself, and this connection was severed by Moon Mode. Per the Rifleman's Creed: \emph{``There are many like it, but this one is mine''}. 

The Moon Mode story is only one extreme example of a generally more subtle tendency of AI to encourage a convergence of forms. Studies find that writers who rely on predictive writing become more predictable and less unique \cite{arnold2020predictive}, and their opinions are influenced by biases in the model \cite{jakesch2023co}. The Moon Mode controversy (and rising discontent against the increasingly opinionated and forceful adjustments made by computational photography more generally \cite{chayka_2022}) signals another cultural reversal. It pushes back against technological convergence of forms and says ``no, I want this to be \emph{my} picture''. The consumer movement to shop local and independent, to consume indie music and cinema, is a countercultural reaction to globalisation and the era of chain stores, supermarkets, and malls. So too might we see consumer movements celebrating independent, artisanal, and craft approaches to knowledge work. This prospect of reversal is one reason to be optimistic that our AI future is not one of institutionalised plagiarism and mindless repetition.

\section{Conclusion}
\label{sec:conclusion}


This position paper began by summarising the dominant view of AI and creativity: that the output of AI systems, in information theoretic terms, cannot be considered creative because of its stochastic reuse nature. We then viewed various alternative conceptions of creativity: as a process, as authorial intent, as interpretation, as reuse, and as randomness -- none of which is amenable to straightforward analyses of information or notation. We discussed the multiple equally valid positions one might adopt on the form-content distinction. We attended to the challenges of relying on intellectual property law for analysing the creativity of AI. Two general patterns were suggested: that creativity is determined contextually by particular communities, and that sources of creativity in a workflow are often determined by the extent to which different portions the workflow are facilitated by mechanical means.

Finally, we looked to contemporary studies of AI-assisted creative knowledge work to glimpse how the loci of creativity are shifting. It appears as though the human labour of \emph{production} is being replaced by AI, while the opportunities for creative human input have shifted to \emph{critical integration}, which is the assessment of AI output and the set of authorial decisions required to incorporate that output into a knowledge workflow. The way in which AI changes our relationship to the production of knowledge work can be seen as analogous to how the industrial revolution changed our relationship to material production. In response we reconfigured our lives and social practices around the new machine-made artefacts, and there was a long, arguably still ongoing period of mutual accommodation between society and industrial technology. It remains to be seen what will emerge from this new transitional phase we find ourselves in. As such, the aim of this paper has been to call for a broader, and more contextually-sensitive attitude to what might constitute creativity in an age of human-AI knowledge work.

\begin{acks}
Thanks to Alan Blackwell for discussions on the topic and reflections on drafts of the paper. Many thanks to my kind reviewers whose suggestions have helped improve the paper. 
\end{acks}

\bibliographystyle{ACM-Reference-Format}
\bibliography{references}


\begin{thebibliography}{148}


\ifx \showCODEN    \undefined \def \showCODEN     #1{\unskip}     \fi
\ifx \showDOI      \undefined \def \showDOI       #1{#1}\fi
\ifx \showISBNx    \undefined \def \showISBNx     #1{\unskip}     \fi
\ifx \showISBNxiii \undefined \def \showISBNxiii  #1{\unskip}     \fi
\ifx \showISSN     \undefined \def \showISSN      #1{\unskip}     \fi
\ifx \showLCCN     \undefined \def \showLCCN      #1{\unskip}     \fi
\ifx \shownote     \undefined \def \shownote      #1{#1}          \fi
\ifx \showarticletitle \undefined \def \showarticletitle #1{#1}   \fi
\ifx \showURL      \undefined \def \showURL       {\relax}        \fi
\providecommand\bibfield[2]{#2}
\providecommand\bibinfo[2]{#2}
\providecommand\natexlab[1]{#1}
\providecommand\showeprint[2][]{arXiv:#2}

\bibitem[Acker et~al\mbox{.}(2011)]%
        {acker2011against}
\bibfield{author}{\bibinfo{person}{Kathy Acker}, \bibinfo{person}{Noah~Eli
  Gordon}, \bibinfo{person}{Peter Gizzi}, \bibinfo{person}{Tan Lin},
  \bibinfo{person}{Trisha Low}, \bibinfo{person}{Juliana Spahr},
  \bibinfo{person}{Bernadette Mayer}, \bibinfo{person}{Harryette Mullen},
  \bibinfo{person}{Claudia Rankine}, {and} \bibinfo{person}{Ron Silliman}.}
  \bibinfo{year}{2011}\natexlab{}.
\newblock \bibinfo{booktitle}{\emph{Against expression: an anthology of
  conceptual writing}}.
\newblock \bibinfo{publisher}{Northwestern University Press}.
\newblock


\bibitem[Adams(2005)]%
        {adams2005tiny}
\bibfield{author}{\bibinfo{person}{Bruce Adams}.}
  \bibinfo{year}{2005}\natexlab{}.
\newblock \bibinfo{booktitle}{\emph{Tiny revolutions in Russia: Twentieth
  century Soviet and Russian history in anecdotes and jokes}}.
\newblock \bibinfo{publisher}{Routledge}.
\newblock


\bibitem[Amihud and Mendelson(1980)]%
        {amihud1980dealership}
\bibfield{author}{\bibinfo{person}{Yakov Amihud} {and} \bibinfo{person}{Haim
  Mendelson}.} \bibinfo{year}{1980}\natexlab{}.
\newblock \showarticletitle{Dealership market: Market-making with inventory}.
\newblock \bibinfo{journal}{\emph{Journal of financial economics}}
  \bibinfo{volume}{8}, \bibinfo{number}{1} (\bibinfo{year}{1980}),
  \bibinfo{pages}{31--53}.
\newblock


\bibitem[Anantrasirichai and Bull(2022)]%
        {anantrasirichai2022artificial}
\bibfield{author}{\bibinfo{person}{Nantheera Anantrasirichai} {and}
  \bibinfo{person}{David Bull}.} \bibinfo{year}{2022}\natexlab{}.
\newblock \showarticletitle{Artificial intelligence in the creative industries:
  a review}.
\newblock \bibinfo{journal}{\emph{Artificial intelligence review}}
  (\bibinfo{year}{2022}), \bibinfo{pages}{1--68}.
\newblock


\bibitem[Angelou(1983)]%
        {angelou1983caged}
\bibfield{author}{\bibinfo{person}{Maya Angelou}.}
  \bibinfo{year}{1983}\natexlab{}.
\newblock \showarticletitle{Caged bird}.
\newblock \bibinfo{journal}{\emph{Shaker, why don't you sing}}
  (\bibinfo{year}{1983}), \bibinfo{pages}{9}.
\newblock


\bibitem[Argyris(1977)]%
        {argyris1977double}
\bibfield{author}{\bibinfo{person}{Chris Argyris}.}
  \bibinfo{year}{1977}\natexlab{}.
\newblock \showarticletitle{Double loop learning in organizations}.
\newblock \bibinfo{journal}{\emph{Harvard business review}}
  \bibinfo{volume}{55}, \bibinfo{number}{5} (\bibinfo{year}{1977}),
  \bibinfo{pages}{115--125}.
\newblock


\bibitem[Arnold et~al\mbox{.}(2020)]%
        {arnold2020predictive}
\bibfield{author}{\bibinfo{person}{Kenneth~C Arnold}, \bibinfo{person}{Krysta
  Chauncey}, {and} \bibinfo{person}{Krzysztof~Z Gajos}.}
  \bibinfo{year}{2020}\natexlab{}.
\newblock \showarticletitle{Predictive text encourages predictable writing}. In
  \bibinfo{booktitle}{\emph{Proceedings of the 25th International Conference on
  Intelligent User Interfaces}}. \bibinfo{pages}{128--138}.
\newblock


\bibitem[Bainbridge(1983)]%
        {bainbridge1983ironies}
\bibfield{author}{\bibinfo{person}{Lisanne Bainbridge}.}
  \bibinfo{year}{1983}\natexlab{}.
\newblock \showarticletitle{Ironies of automation}.
\newblock In \bibinfo{booktitle}{\emph{Analysis, design and evaluation of
  man--machine systems}}. \bibinfo{publisher}{Elsevier},
  \bibinfo{pages}{129--135}.
\newblock


\bibitem[Barthes(2016)]%
        {barthes2016death}
\bibfield{author}{\bibinfo{person}{Roland Barthes}.}
  \bibinfo{year}{2016}\natexlab{}.
\newblock \showarticletitle{The death of the author}.
\newblock In \bibinfo{booktitle}{\emph{Readings in the Theory of Religion}}.
  \bibinfo{publisher}{Routledge}, \bibinfo{pages}{141--145}.
\newblock


\bibitem[Bender et~al\mbox{.}(2021)]%
        {bender2021dangers}
\bibfield{author}{\bibinfo{person}{Emily~M Bender}, \bibinfo{person}{Timnit
  Gebru}, \bibinfo{person}{Angelina McMillan-Major}, {and}
  \bibinfo{person}{Shmargaret Shmitchell}.} \bibinfo{year}{2021}\natexlab{}.
\newblock \showarticletitle{On the Dangers of Stochastic Parrots: Can Language
  Models Be Too Big?}. In \bibinfo{booktitle}{\emph{Proceedings of the 2021 ACM
  conference on fairness, accountability, and transparency}}.
  \bibinfo{pages}{610--623}.
\newblock


\bibitem[Benjamin(1935)]%
        {benjamin1935work}
\bibfield{author}{\bibinfo{person}{Walter Benjamin}.}
  \bibinfo{year}{1935}\natexlab{}.
\newblock \bibinfo{title}{The Work of Art in the Age of Mechanical
  Reproduction, 1936}.
\newblock
\newblock


\bibitem[Benjamin and Rendall(2021)]%
        {benjamin2021translator}
\bibfield{author}{\bibinfo{person}{Walter Benjamin} {and}
  \bibinfo{person}{Steven Rendall}.} \bibinfo{year}{2021}\natexlab{}.
\newblock \showarticletitle{The translator’s task}.
\newblock In \bibinfo{booktitle}{\emph{The translation studies reader}}.
  \bibinfo{publisher}{Routledge}, \bibinfo{pages}{89--97}.
\newblock


\bibitem[Bently(1994)]%
        {bently1994copyright}
\bibfield{author}{\bibinfo{person}{Lionel Bently}.}
  \bibinfo{year}{1994}\natexlab{}.
\newblock \showarticletitle{Copyright and the Death of the Author in Literature
  and Law}.
\newblock \bibinfo{journal}{\emph{Mod. L. Rev.}}  \bibinfo{volume}{57}
  (\bibinfo{year}{1994}), \bibinfo{pages}{973}.
\newblock


\bibitem[Bernard(1993)]%
        {bernard1993minimalist}
\bibfield{author}{\bibinfo{person}{Jonathan~W Bernard}.}
  \bibinfo{year}{1993}\natexlab{}.
\newblock \showarticletitle{The minimalist aesthetic in the plastic arts and in
  music}.
\newblock \bibinfo{journal}{\emph{Perspectives of New Music}}
  (\bibinfo{year}{1993}), \bibinfo{pages}{86--132}.
\newblock


\bibitem[Blackwell(2020)]%
        {blackwell2020objective}
\bibfield{author}{\bibinfo{person}{Alan~F Blackwell}.}
  \bibinfo{year}{2020}\natexlab{}.
\newblock \showarticletitle{Objective functions:(In) humanity and inequity in
  artificial intelligence}.
\newblock \bibinfo{journal}{\emph{Science in the ForeSt, Science in the PaSt}}
  (\bibinfo{year}{2020}), \bibinfo{pages}{191}.
\newblock


\bibitem[Blackwell(2022)]%
        {blackwell2022coding}
\bibfield{author}{\bibinfo{person}{Alan~F. Blackwell}.}
  \bibinfo{year}{2022}\natexlab{}.
\newblock \showarticletitle{Coding or AI? Tools for Control, Surprise and
  Creativity}. In \bibinfo{booktitle}{\emph{{Proceedings of the 33rd Annual
  Conference of the Psychology of Programming Interest Group (PPIG 2022)}}}.
\newblock


\bibitem[Bloom et~al\mbox{.}(1997)]%
        {bloom1997anxiety}
\bibfield{author}{\bibinfo{person}{Harold Bloom} {et~al\mbox{.}}}
  \bibinfo{year}{1997}\natexlab{}.
\newblock \bibinfo{booktitle}{\emph{The anxiety of influence: A theory of
  poetry}}.
\newblock \bibinfo{publisher}{Oxford University Press, USA}.
\newblock


\bibitem[Boden(2008)]%
        {boden2008computers}
\bibfield{author}{\bibinfo{person}{Margaret Boden}.}
  \bibinfo{year}{2008}\natexlab{}.
\newblock \showarticletitle{Computers and creativity: models and applications}.
\newblock \bibinfo{journal}{\emph{The Routledge Companion to Creativity}}
  (\bibinfo{year}{2008}), \bibinfo{pages}{179--188}.
\newblock


\bibitem[Boden(2007)]%
        {boden_2007}
\bibfield{author}{\bibinfo{person}{Margaret~A. Boden}.}
  \bibinfo{year}{2007}\natexlab{}.
\newblock \showarticletitle{Creativity in a nutshell}.
\newblock \bibinfo{journal}{\emph{Think}} \bibinfo{volume}{5},
  \bibinfo{number}{15} (\bibinfo{year}{2007}), \bibinfo{pages}{83–96}.
\newblock
\urldef\tempurl%
\url{https://doi.org/10.1017/S147717560000230X}
\showDOI{\tempurl}


\bibitem[Boden and Edmonds(2009)]%
        {boden2009generative}
\bibfield{author}{\bibinfo{person}{Margaret~A Boden} {and}
  \bibinfo{person}{Ernest~A Edmonds}.} \bibinfo{year}{2009}\natexlab{}.
\newblock \showarticletitle{What is generative art?}
\newblock \bibinfo{journal}{\emph{Digital Creativity}} \bibinfo{volume}{20},
  \bibinfo{number}{1-2} (\bibinfo{year}{2009}), \bibinfo{pages}{21--46}.
\newblock


\bibitem[Bommasani et~al\mbox{.}(2021)]%
        {bommasani2021opportunities}
\bibfield{author}{\bibinfo{person}{Rishi Bommasani}, \bibinfo{person}{Drew~A
  Hudson}, \bibinfo{person}{Ehsan Adeli}, \bibinfo{person}{Russ Altman},
  \bibinfo{person}{Simran Arora}, \bibinfo{person}{Sydney von Arx},
  \bibinfo{person}{Michael~S Bernstein}, \bibinfo{person}{Jeannette Bohg},
  \bibinfo{person}{Antoine Bosselut}, \bibinfo{person}{Emma Brunskill},
  {et~al\mbox{.}}} \bibinfo{year}{2021}\natexlab{}.
\newblock \showarticletitle{On the opportunities and risks of foundation
  models}.
\newblock \bibinfo{journal}{\emph{arXiv preprint arXiv:2108.07258}}
  (\bibinfo{year}{2021}).
\newblock


\bibitem[Borges(1962)]%
        {borges1962ficciones}
\bibfield{author}{\bibinfo{person}{Jorge~Luis Borges}.}
  \bibinfo{year}{1962}\natexlab{}.
\newblock \bibinfo{booktitle}{\emph{Ficciones}}. Vol.~\bibinfo{volume}{320}.
\newblock \bibinfo{publisher}{Grove Press}.
\newblock


\bibitem[Braverman(1998)]%
        {braverman1998labor}
\bibfield{author}{\bibinfo{person}{Harry Braverman}.}
  \bibinfo{year}{1998}\natexlab{}.
\newblock \bibinfo{booktitle}{\emph{Labor and monopoly capital: The degradation
  of work in the twentieth century}}.
\newblock \bibinfo{publisher}{NYU Press}.
\newblock


\bibitem[Brengelman(1980)]%
        {brengelman1980orthoepists}
\bibfield{author}{\bibinfo{person}{Fred~H Brengelman}.}
  \bibinfo{year}{1980}\natexlab{}.
\newblock \showarticletitle{Orthoepists, printers, and the rationalization of
  English spelling}.
\newblock \bibinfo{journal}{\emph{The Journal of English and Germanic
  Philology}} \bibinfo{volume}{79}, \bibinfo{number}{3} (\bibinfo{year}{1980}),
  \bibinfo{pages}{332--354}.
\newblock


\bibitem[Brown(2019)]%
        {brown_2019}
\bibfield{author}{\bibinfo{person}{C.~Scott Brown}.}
  \bibinfo{year}{2019}\natexlab{}.
\newblock \bibinfo{title}{Huawei P30 Pro 'Moon Mode' stirs controversy (update:
  Huawei responds)}.
\newblock
\newblock
\urldef\tempurl%
\url{https://www.androidauthority.com/huawei-p30-pro-moon-mode-controversy-978486/}
\showURL{%
\tempurl}


\bibitem[Brown et~al\mbox{.}(2020)]%
        {brown2020language}
\bibfield{author}{\bibinfo{person}{Tom Brown}, \bibinfo{person}{Benjamin Mann},
  \bibinfo{person}{Nick Ryder}, \bibinfo{person}{Melanie Subbiah},
  \bibinfo{person}{Jared~D Kaplan}, \bibinfo{person}{Prafulla Dhariwal},
  \bibinfo{person}{Arvind Neelakantan}, \bibinfo{person}{Pranav Shyam},
  \bibinfo{person}{Girish Sastry}, \bibinfo{person}{Amanda Askell},
  {et~al\mbox{.}}} \bibinfo{year}{2020}\natexlab{}.
\newblock \showarticletitle{Language models are few-shot learners}.
\newblock \bibinfo{journal}{\emph{Advances in neural information processing
  systems}}  \bibinfo{volume}{33} (\bibinfo{year}{2020}),
  \bibinfo{pages}{1877--1901}.
\newblock


\bibitem[Bryson et~al\mbox{.}(2017)]%
        {bryson2017and}
\bibfield{author}{\bibinfo{person}{Joanna~J Bryson},
  \bibinfo{person}{Mihailis~E Diamantis}, {and} \bibinfo{person}{Thomas~D
  Grant}.} \bibinfo{year}{2017}\natexlab{}.
\newblock \showarticletitle{Of, for, and by the people: the legal lacuna of
  synthetic persons}.
\newblock \bibinfo{journal}{\emph{Artificial Intelligence and Law}}
  \bibinfo{volume}{25} (\bibinfo{year}{2017}), \bibinfo{pages}{273--291}.
\newblock


\bibitem[Bush et~al\mbox{.}(1945)]%
        {bush1945we}
\bibfield{author}{\bibinfo{person}{Vannevar Bush} {et~al\mbox{.}}}
  \bibinfo{year}{1945}\natexlab{}.
\newblock \showarticletitle{As we may think}.
\newblock \bibinfo{journal}{\emph{The atlantic monthly}} \bibinfo{volume}{176},
  \bibinfo{number}{1} (\bibinfo{year}{1945}), \bibinfo{pages}{101--108}.
\newblock


\bibitem[Cantor(2022)]%
        {cantor_2022}
\bibfield{author}{\bibinfo{person}{Paul Cantor}.}
  \bibinfo{year}{2022}\natexlab{}.
\newblock \bibinfo{title}{Did Beyonce really sample Robin S "Show me love" on
  New Song "Break My Soul"?}
\newblock
\newblock
\urldef\tempurl%
\url{https://paulcantor.medium.com/did-beyonce-really-sample-robin-s-show-me-love-on-new-song-break-my-soul-9cbf1d4b626a}
\showURL{%
\tempurl}


\bibitem[Caramiaux and Fdili~Alaoui(2022)]%
        {caramiaux2022explorers}
\bibfield{author}{\bibinfo{person}{Baptiste Caramiaux} {and}
  \bibinfo{person}{Sarah Fdili~Alaoui}.} \bibinfo{year}{2022}\natexlab{}.
\newblock \showarticletitle{" Explorers of Unknown Planets" Practices and
  Politics of Artificial Intelligence in Visual Arts}.
\newblock \bibinfo{journal}{\emph{Proceedings of the ACM on Human-Computer
  Interaction}} \bibinfo{volume}{6}, \bibinfo{number}{CSCW2}
  (\bibinfo{year}{2022}), \bibinfo{pages}{1--24}.
\newblock


\bibitem[Chayka(2022)]%
        {chayka_2022}
\bibfield{author}{\bibinfo{person}{Kyle Chayka}.}
  \bibinfo{year}{2022}\natexlab{}.
\newblock \bibinfo{title}{Have iphone cameras become too smart?}
\newblock
\newblock
\urldef\tempurl%
\url{https://www.newyorker.com/culture/infinite-scroll/have-iphone-cameras-become-too-smart}
\showURL{%
\tempurl}


\bibitem[Che and Chien(2022)]%
        {che_chien_2022}
\bibfield{author}{\bibinfo{person}{Chang Che} {and} \bibinfo{person}{Amy~Chang
  Chien}.} \bibinfo{year}{2022}\natexlab{}.
\newblock \bibinfo{title}{Memes, puns and blank sheets of paper: China's
  creative acts of protest}.
\newblock
\newblock
\urldef\tempurl%
\url{https://www.nytimes.com/2022/11/28/world/asia/china-protests-blank-sheets.html}
\showURL{%
\tempurl}


\bibitem[Chung et~al\mbox{.}(2022)]%
        {chung2022artist}
\bibfield{author}{\bibinfo{person}{John Joon~Young Chung},
  \bibinfo{person}{Shiqing He}, {and} \bibinfo{person}{Eytan Adar}.}
  \bibinfo{year}{2022}\natexlab{}.
\newblock \showarticletitle{Artist support networks: Implications for future
  creativity support tools}. In \bibinfo{booktitle}{\emph{Designing Interactive
  Systems Conference}}. \bibinfo{pages}{232--246}.
\newblock


\bibitem[Clement(2018)]%
        {clement2018brutalism}
\bibfield{author}{\bibinfo{person}{Alexander Clement}.}
  \bibinfo{year}{2018}\natexlab{}.
\newblock \bibinfo{booktitle}{\emph{Brutalism: post-war British architecture}}.
\newblock \bibinfo{publisher}{The Crowood Press}.
\newblock


\bibitem[Collins et~al\mbox{.}(1998)]%
        {collins1998shape}
\bibfield{author}{\bibinfo{person}{Harry~M Collins}, \bibinfo{person}{Martin
  Kusch}, {et~al\mbox{.}}} \bibinfo{year}{1998}\natexlab{}.
\newblock \bibinfo{booktitle}{\emph{The shape of actions: What humans and
  machines can do}}.
\newblock \bibinfo{publisher}{MIT press}.
\newblock


\bibitem[Constantine(2019)]%
        {constantine2019possibilists}
\bibfield{author}{\bibinfo{person}{Mary-Ann Constantine}.}
  \bibinfo{year}{2019}\natexlab{}.
\newblock \showarticletitle{{79The possibilists: Romantic-era literary forgery
  and British alternative pasts}}.
\newblock In \bibinfo{booktitle}{\emph{{Counterfactual Romanticism}}}.
  \bibinfo{publisher}{Manchester University Press}.
\newblock
\showISBNx{9781784991418}
\urldef\tempurl%
\url{https://doi.org/10.7765/9781526107077.00009}
\showDOI{\tempurl}
\showeprint{https://academic.oup.com/manchester-scholarship-online/book/0/chapter/262463140/chapter-ag-pdf/44552686/book\_30839\_section\_262463140.ag.pdf}


\bibitem[Crawford(2018)]%
        {crawford_2018}
\bibfield{author}{\bibinfo{person}{Jason Crawford}.}
  \bibinfo{year}{2018}\natexlab{}.
\newblock \bibinfo{title}{Out of whole cloth}.
\newblock
\newblock
\urldef\tempurl%
\url{https://rootsofprogress.org/out-of-whole-cloth}
\showURL{%
\tempurl}


\bibitem[Currie and Allart(2012)]%
        {currie2012pieter}
\bibfield{author}{\bibinfo{person}{Christina Currie} {and}
  \bibinfo{person}{Dominique Allart}.} \bibinfo{year}{2012}\natexlab{}.
\newblock \showarticletitle{Pieter Brueghel as a copyist after Pieter Bruegel}.
\newblock  (\bibinfo{year}{2012}).
\newblock


\bibitem[Dang et~al\mbox{.}(2022)]%
        {dang2022beyond}
\bibfield{author}{\bibinfo{person}{Hai Dang}, \bibinfo{person}{Karim
  Benharrak}, \bibinfo{person}{Florian Lehmann}, {and} \bibinfo{person}{Daniel
  Buschek}.} \bibinfo{year}{2022}\natexlab{}.
\newblock \showarticletitle{Beyond Text Generation: Supporting Writers with
  Continuous Automatic Text Summaries}. In
  \bibinfo{booktitle}{\emph{Proceedings of the 35th Annual ACM Symposium on
  User Interface Software and Technology}}. \bibinfo{pages}{1--13}.
\newblock


\bibitem[Davis(2018)]%
        {davis20189th}
\bibfield{author}{\bibinfo{person}{Jordan Davis}.}
  \bibinfo{year}{2018}\natexlab{}.
\newblock \showarticletitle{9th Circuit Orders Retrial Over “Stairway to
  Heaven” Copyright Infringement Case}.
\newblock  (\bibinfo{year}{2018}).
\newblock


\bibitem[Day(1966)]%
        {day1966locke}
\bibfield{author}{\bibinfo{person}{John~P Day}.}
  \bibinfo{year}{1966}\natexlab{}.
\newblock \showarticletitle{Locke on property}.
\newblock \bibinfo{journal}{\emph{The Philosophical Quarterly (1950-)}}
  \bibinfo{volume}{16}, \bibinfo{number}{64} (\bibinfo{year}{1966}),
  \bibinfo{pages}{207--220}.
\newblock


\bibitem[Debord and Wolman(1956)]%
        {debord1956methods}
\bibfield{author}{\bibinfo{person}{Guy Debord} {and} \bibinfo{person}{Gil~J
  Wolman}.} \bibinfo{year}{1956}\natexlab{}.
\newblock \showarticletitle{Methods of detournement}.
\newblock \bibinfo{journal}{\emph{Les L{\`e}vres Nues}}  \bibinfo{volume}{8}
  (\bibinfo{year}{1956}).
\newblock


\bibitem[Deck(2022)]%
        {deck_2022}
\bibfield{author}{\bibinfo{person}{Andrew Deck}.}
  \bibinfo{year}{2022}\natexlab{}.
\newblock \bibinfo{title}{Ai-generated art sparks furious backlash from Japan's
  Anime Community}.
\newblock
\newblock
\urldef\tempurl%
\url{https://restofworld.org/2022/ai-backlash-anime-artists/}
\showURL{%
\tempurl}


\bibitem[Dennett(1971)]%
        {dennett1971intentional}
\bibfield{author}{\bibinfo{person}{Daniel~C Dennett}.}
  \bibinfo{year}{1971}\natexlab{}.
\newblock \showarticletitle{Intentional systems}.
\newblock \bibinfo{journal}{\emph{The Journal of Philosophy}}
  \bibinfo{volume}{68}, \bibinfo{number}{4} (\bibinfo{year}{1971}),
  \bibinfo{pages}{87--106}.
\newblock


\bibitem[Draper(1971)]%
        {draper1971concrete}
\bibfield{author}{\bibinfo{person}{Ronald~P Draper}.}
  \bibinfo{year}{1971}\natexlab{}.
\newblock \showarticletitle{Concrete poetry}.
\newblock \bibinfo{journal}{\emph{New Literary History}} \bibinfo{volume}{2},
  \bibinfo{number}{2} (\bibinfo{year}{1971}), \bibinfo{pages}{329--340}.
\newblock


\bibitem[Drucker(1959)]%
        {drucker1959}
\bibfield{author}{\bibinfo{person}{Peter~F. Drucker}.}
  \bibinfo{year}{1959}\natexlab{}.
\newblock \bibinfo{booktitle}{\emph{Landmarks of Tomorrow}
  (\bibinfo{edition}{1st ed.} ed.)}.
\newblock \bibinfo{publisher}{Harper}, \bibinfo{address}{New York}.
\newblock


\bibitem[Duggan(2022)]%
        {duggan2022transformative}
\bibfield{author}{\bibinfo{person}{Jennifer Duggan}.}
  \bibinfo{year}{2022}\natexlab{}.
\newblock \showarticletitle{Transformative readings: Harry Potter fan fiction,
  trans/queer reader response, and JK Rowling}.
\newblock \bibinfo{journal}{\emph{Children's Literature in Education}}
  \bibinfo{volume}{53}, \bibinfo{number}{2} (\bibinfo{year}{2022}),
  \bibinfo{pages}{147--168}.
\newblock


\bibitem[Edwards(2022)]%
        {edwards_2022}
\bibfield{author}{\bibinfo{person}{Benj Edwards}.}
  \bibinfo{year}{2022}\natexlab{}.
\newblock \bibinfo{title}{DeviantArt upsets artists with its new AI Art
  Generator, DreamUp}.
\newblock
\newblock
\urldef\tempurl%
\url{https://arstechnica.com/information-technology/2022/11/deviantart-upsets-artists-with-its-new-ai-art-generator-dreamup/}
\showURL{%
\tempurl}


\bibitem[Eno and Schmidt(1975)]%
        {eno1975oblique}
\bibfield{author}{\bibinfo{person}{Brian Eno} {and} \bibinfo{person}{Peter
  Schmidt}.} \bibinfo{year}{1975}\natexlab{}.
\newblock \showarticletitle{Oblique strategies}.
\newblock \bibinfo{journal}{\emph{Opal.(Limited edition, boxed set of
  cards.)[rMAB]}} (\bibinfo{year}{1975}).
\newblock


\bibitem[Ferguson(2022)]%
        {ferguson_2022}
\bibfield{author}{\bibinfo{person}{Kirby Ferguson}.}
  \bibinfo{year}{2022}\natexlab{}.
\newblock \bibinfo{title}{Everything is a remix}.
\newblock
\newblock
\urldef\tempurl%
\url{https://www.everythingisaremix.info/}
\showURL{%
\tempurl}


\bibitem[Foucault(2017)]%
        {foucault2017author}
\bibfield{author}{\bibinfo{person}{Michel Foucault}.}
  \bibinfo{year}{2017}\natexlab{}.
\newblock \showarticletitle{What is an Author?}
\newblock In \bibinfo{booktitle}{\emph{Aesthetics}}.
  \bibinfo{publisher}{Routledge}, \bibinfo{pages}{284--288}.
\newblock


\bibitem[Frich et~al\mbox{.}(2018)]%
        {frich2018twenty}
\bibfield{author}{\bibinfo{person}{Jonas Frich}, \bibinfo{person}{Michael
  Mose~Biskjaer}, {and} \bibinfo{person}{Peter Dalsgaard}.}
  \bibinfo{year}{2018}\natexlab{}.
\newblock \showarticletitle{Twenty years of creativity research in
  human-computer interaction: Current state and future directions}. In
  \bibinfo{booktitle}{\emph{Proceedings of the 2018 Designing Interactive
  Systems Conference}}. \bibinfo{pages}{1235--1257}.
\newblock


\bibitem[Froman(1988)]%
        {froman1988action}
\bibfield{author}{\bibinfo{person}{Wayne~J Froman}.}
  \bibinfo{year}{1988}\natexlab{}.
\newblock \showarticletitle{Action painting and the world-as-picture}.
\newblock \bibinfo{journal}{\emph{The Journal of aesthetics and art criticism}}
  \bibinfo{volume}{46}, \bibinfo{number}{4} (\bibinfo{year}{1988}),
  \bibinfo{pages}{469--475}.
\newblock


\bibitem[Galenson(2011)]%
        {galenson2011old}
\bibfield{author}{\bibinfo{person}{David~W Galenson}.}
  \bibinfo{year}{2011}\natexlab{}.
\newblock \showarticletitle{Old masters and young geniuses}.
\newblock In \bibinfo{booktitle}{\emph{Old Masters and Young Geniuses}}.
  \bibinfo{publisher}{Princeton University Press}.
\newblock


\bibitem[Gangopadhyay and Mondal(2012)]%
        {gangopadhyay2012does}
\bibfield{author}{\bibinfo{person}{Kausik Gangopadhyay} {and}
  \bibinfo{person}{Debasis Mondal}.} \bibinfo{year}{2012}\natexlab{}.
\newblock \showarticletitle{Does stronger protection of intellectual property
  stimulate innovation?}
\newblock \bibinfo{journal}{\emph{Economics Letters}} \bibinfo{volume}{116},
  \bibinfo{number}{1} (\bibinfo{year}{2012}), \bibinfo{pages}{80--82}.
\newblock


\bibitem[Gaver(2012)]%
        {gaver2012should}
\bibfield{author}{\bibinfo{person}{William Gaver}.}
  \bibinfo{year}{2012}\natexlab{}.
\newblock \showarticletitle{What should we expect from research through
  design?}. In \bibinfo{booktitle}{\emph{Proceedings of the SIGCHI conference
  on human factors in computing systems}}. \bibinfo{pages}{937--946}.
\newblock


\bibitem[Gero et~al\mbox{.}(2022)]%
        {gero2022sparks}
\bibfield{author}{\bibinfo{person}{Katy~Ilonka Gero}, \bibinfo{person}{Vivian
  Liu}, {and} \bibinfo{person}{Lydia Chilton}.}
  \bibinfo{year}{2022}\natexlab{}.
\newblock \showarticletitle{Sparks: Inspiration for science writing using
  language models}. In \bibinfo{booktitle}{\emph{Designing Interactive Systems
  Conference}}. \bibinfo{pages}{1002--1019}.
\newblock


\bibitem[Ginsburg(1992)]%
        {ginsburg1992no}
\bibfield{author}{\bibinfo{person}{Jane~C Ginsburg}.}
  \bibinfo{year}{1992}\natexlab{}.
\newblock \showarticletitle{No Sweat Copyright and Other Protection of Works of
  Information after Feist v. Rural Telephone}.
\newblock \bibinfo{journal}{\emph{Colum. L. Rev.}}  \bibinfo{volume}{92}
  (\bibinfo{year}{1992}), \bibinfo{pages}{338}.
\newblock


\bibitem[Goodman(2008)]%
        {goodman2008speaking}
\bibfield{author}{\bibinfo{person}{Felicitas~D Goodman}.}
  \bibinfo{year}{2008}\natexlab{}.
\newblock \bibinfo{booktitle}{\emph{Speaking in tongues: A cross-cultural study
  of glossolalia}}.
\newblock \bibinfo{publisher}{Wipf and Stock Publishers}.
\newblock


\bibitem[Grace(2004)]%
        {grace2004effect}
\bibfield{author}{\bibinfo{person}{Cheri Grace}.}
  \bibinfo{year}{2004}\natexlab{}.
\newblock \showarticletitle{The effect of changing intellectual property on
  pharmaceutical industry prospects in India and China}.
\newblock \bibinfo{journal}{\emph{DFID Health Systems Resource Centre}}
  (\bibinfo{year}{2004}), \bibinfo{pages}{1--68}.
\newblock


\bibitem[Greenbaum(1996)]%
        {greenbaum1996back}
\bibfield{author}{\bibinfo{person}{Joan Greenbaum}.}
  \bibinfo{year}{1996}\natexlab{}.
\newblock \showarticletitle{Back to Labor: Returning to labor process
  discussions in the study of work}. In \bibinfo{booktitle}{\emph{Proceedings
  of the 1996 ACM conference on Computer supported cooperative work}}.
  \bibinfo{pages}{229--237}.
\newblock


\bibitem[Grice(1975)]%
        {grice1975logic}
\bibfield{author}{\bibinfo{person}{Herbert~P Grice}.}
  \bibinfo{year}{1975}\natexlab{}.
\newblock \showarticletitle{Logic and conversation}.
\newblock In \bibinfo{booktitle}{\emph{Speech acts}}.
  \bibinfo{publisher}{Brill}, \bibinfo{pages}{41--58}.
\newblock


\bibitem[Groot(1964)]%
        {groot1964religious}
\bibfield{author}{\bibinfo{person}{Jan Jakob~Maria Groot}.}
  \bibinfo{year}{1964}\natexlab{}.
\newblock \bibinfo{booktitle}{\emph{The religious system of China}}.
  Vol.~\bibinfo{volume}{4}.
\newblock \bibinfo{publisher}{Brill Archive}.
\newblock


\bibitem[Harding(2022)]%
        {harding_2022}
\bibfield{author}{\bibinfo{person}{Charlie Harding}.}
  \bibinfo{year}{2022}\natexlab{}.
\newblock \bibinfo{title}{Beyoncé's House}.
\newblock
\newblock
\urldef\tempurl%
\url{https://switchedonpop.com/episodes/beyonce-house-break-my-soul-show-me-love}
\showURL{%
\tempurl}


\bibitem[Hayes and Youderian(2013)]%
        {hayes2013ultimate}
\bibfield{author}{\bibinfo{person}{Mark Hayes} {and} \bibinfo{person}{Andrew
  Youderian}.} \bibinfo{year}{2013}\natexlab{}.
\newblock \bibinfo{booktitle}{\emph{The ultimate guide to dropshipping}}.
\newblock \bibinfo{publisher}{Lulu. com}.
\newblock


\bibitem[Hess(2006)]%
        {hess2006foucault}
\bibfield{author}{\bibinfo{person}{Mickey Hess}.}
  \bibinfo{year}{2006}\natexlab{}.
\newblock \showarticletitle{Was Foucault a plagiarist? Hip-hop sampling and
  academic citation}.
\newblock \bibinfo{journal}{\emph{Computers and Composition}}
  \bibinfo{volume}{23}, \bibinfo{number}{3} (\bibinfo{year}{2006}),
  \bibinfo{pages}{280--295}.
\newblock


\bibitem[Higgins(2004)]%
        {higgins_2004}
\bibfield{author}{\bibinfo{person}{Charlotte Higgins}.}
  \bibinfo{year}{2004}\natexlab{}.
\newblock \bibinfo{title}{Work of art that inspired a movement ... a urinal}.
\newblock
\newblock
\urldef\tempurl%
\url{https://www.theguardian.com/uk/2004/dec/02/arts.artsnews1}
\showURL{%
\tempurl}


\bibitem[Howard-Hill(2006)]%
        {howard2006early}
\bibfield{author}{\bibinfo{person}{Trevor~Howard Howard-Hill}.}
  \bibinfo{year}{2006}\natexlab{}.
\newblock \showarticletitle{Early modern printers and the standardization of
  English spelling}.
\newblock \bibinfo{journal}{\emph{Modern Language Review}}
  \bibinfo{volume}{101}, \bibinfo{number}{1} (\bibinfo{year}{2006}),
  \bibinfo{pages}{16--29}.
\newblock


\bibitem[Inie et~al\mbox{.}(2023)]%
        {inie2023designing}
\bibfield{author}{\bibinfo{person}{Nanna Inie}, \bibinfo{person}{Jeanette
  Falk}, {and} \bibinfo{person}{Steven Tanimoto}.}
  \bibinfo{year}{2023}\natexlab{}.
\newblock \showarticletitle{Designing Participatory AI: Creative Professionals'
  Worries and Expectations about Generative AI}.
\newblock \bibinfo{journal}{\emph{arXiv preprint arXiv:2303.08931}}
  (\bibinfo{year}{2023}).
\newblock


\bibitem[Irwin(2015)]%
        {irwin2015authorial}
\bibfield{author}{\bibinfo{person}{William Irwin}.}
  \bibinfo{year}{2015}\natexlab{}.
\newblock \showarticletitle{Authorial declaration and extreme actual
  intentionalism: Is Dumbledore gay?}
\newblock \bibinfo{journal}{\emph{The Journal of Aesthetics and Art Criticism}}
  \bibinfo{volume}{73}, \bibinfo{number}{2} (\bibinfo{year}{2015}),
  \bibinfo{pages}{141--147}.
\newblock


\bibitem[Iser and Tompkins(1984)]%
        {iser1984reader}
\bibfield{author}{\bibinfo{person}{Wolfgang Iser} {and} \bibinfo{person}{Jane
  Tompkins}.} \bibinfo{year}{1984}\natexlab{}.
\newblock \bibinfo{booktitle}{\emph{Reader-response criticism}}.
\newblock


\bibitem[Jakesch et~al\mbox{.}(2023)]%
        {jakesch2023co}
\bibfield{author}{\bibinfo{person}{Maurice Jakesch}, \bibinfo{person}{Advait
  Bhat}, \bibinfo{person}{Daniel Buschek}, \bibinfo{person}{Lior Zalmanson},
  {and} \bibinfo{person}{Mor Naaman}.} \bibinfo{year}{2023}\natexlab{}.
\newblock \showarticletitle{Co-Writing with Opinionated Language Models Affects
  Users' Views}.
\newblock \bibinfo{journal}{\emph{arXiv preprint arXiv:2302.00560}}
  (\bibinfo{year}{2023}).
\newblock


\bibitem[James(2009)]%
        {james2009constraining}
\bibfield{author}{\bibinfo{person}{Alison James}.}
  \bibinfo{year}{2009}\natexlab{}.
\newblock \bibinfo{booktitle}{\emph{Constraining Chance: Georges Perec and the
  Oulipo}}.
\newblock \bibinfo{publisher}{Northwestern University Press}.
\newblock


\bibitem[Jenkins(2014)]%
        {jenkins2014last}
\bibfield{author}{\bibinfo{person}{Jennifer Jenkins}.}
  \bibinfo{year}{2014}\natexlab{}.
\newblock \showarticletitle{Last sale?: Libraries’ rights in the digital
  age}.
\newblock \bibinfo{journal}{\emph{College \& Research Libraries News}}
  \bibinfo{volume}{75}, \bibinfo{number}{2} (\bibinfo{year}{2014}),
  \bibinfo{pages}{69--75}.
\newblock


\bibitem[Juzwiak(2022)]%
        {juzwiak_2022}
\bibfield{author}{\bibinfo{person}{Rich Juzwiak}.}
  \bibinfo{year}{2022}\natexlab{}.
\newblock \bibinfo{title}{Beyoncé's 'Break My Soul' and the Long Tail of 'Show
  Me Love'}.
\newblock
\newblock
\urldef\tempurl%
\url{https://www.nytimes.com/2022/06/27/arts/music/beyonce-break-my-soul-robin-s-show-me-love.html}
\showURL{%
\tempurl}


\bibitem[Karakilic(2022)]%
        {karakilic2022humans}
\bibfield{author}{\bibinfo{person}{Emrah Karakilic}.}
  \bibinfo{year}{2022}\natexlab{}.
\newblock \showarticletitle{Why do humans remain central to the knowledge work
  in the age of robots? Marx’s Fragment on machines and beyond}.
\newblock \bibinfo{journal}{\emph{Work, Employment and Society}}
  \bibinfo{volume}{36}, \bibinfo{number}{1} (\bibinfo{year}{2022}),
  \bibinfo{pages}{179--189}.
\newblock


\bibitem[Katz(2012)]%
        {katz2012groove}
\bibfield{author}{\bibinfo{person}{Mark Katz}.}
  \bibinfo{year}{2012}\natexlab{}.
\newblock \bibinfo{booktitle}{\emph{Groove music: The art and culture of the
  hip-hop DJ}}.
\newblock \bibinfo{publisher}{Oxford University Press on Demand}.
\newblock


\bibitem[Kidd(1994)]%
        {kidd1994marks}
\bibfield{author}{\bibinfo{person}{Alison Kidd}.}
  \bibinfo{year}{1994}\natexlab{}.
\newblock \showarticletitle{The marks are on the knowledge worker}. In
  \bibinfo{booktitle}{\emph{Proceedings of the SIGCHI conference on Human
  factors in computing systems}}. \bibinfo{pages}{186--191}.
\newblock


\bibitem[Kopenkina(2022)]%
        {kopenkina_2022}
\bibfield{author}{\bibinfo{person}{Oksana Kopenkina}.}
  \bibinfo{year}{2022}\natexlab{}.
\newblock \bibinfo{title}{Pieter Bruegel the younger. copyist or great artist?
  - arts diary \& PAD}.
\newblock
\newblock
\urldef\tempurl%
\url{https://arts-pad.com/pieter-bruegel-the-younger/}
\showURL{%
\tempurl}


\bibitem[Latham(2003)]%
        {latham2003newton}
\bibfield{author}{\bibinfo{person}{Susan~J Latham}.}
  \bibinfo{year}{2003}\natexlab{}.
\newblock \showarticletitle{Newton v. Diamond; Measuring the Legitimacy of
  Unauthorized Compositional Samplling-A Clue Illuminated and Obscured}.
\newblock \bibinfo{journal}{\emph{Hastings Comm. \& Ent. LJ}}
  \bibinfo{volume}{26} (\bibinfo{year}{2003}), \bibinfo{pages}{119}.
\newblock


\bibitem[Learning(2023)]%
        {learning_2023}
\bibfield{author}{\bibinfo{person}{MoMA Learning}.}
  \bibinfo{year}{2023}\natexlab{}.
\newblock \bibinfo{title}{Marcel Duchamp and the Readymade}.
\newblock
\newblock
\urldef\tempurl%
\url{https://www.moma.org/learn/moma_learning/themes/dada/marcel-duchamp-and-the-readymade/}
\showURL{%
\tempurl}


\bibitem[Lee et~al\mbox{.}(2022)]%
        {lee2022language}
\bibfield{author}{\bibinfo{person}{Jooyoung Lee}, \bibinfo{person}{Thai Le},
  \bibinfo{person}{Jinghui Chen}, {and} \bibinfo{person}{Dongwon Lee}.}
  \bibinfo{year}{2022}\natexlab{}.
\newblock \showarticletitle{Do Language Models Plagiarize?}
\newblock \bibinfo{journal}{\emph{arXiv preprint arXiv:2203.07618}}
  (\bibinfo{year}{2022}).
\newblock


\bibitem[Lee(2018)]%
        {lee2018ai}
\bibfield{author}{\bibinfo{person}{Kai-Fu Lee}.}
  \bibinfo{year}{2018}\natexlab{}.
\newblock \bibinfo{booktitle}{\emph{AI superpowers: China, Silicon Valley, and
  the new world order}}.
\newblock \bibinfo{publisher}{Houghton Mifflin}.
\newblock


\bibitem[Lemley(2009)]%
        {lemley2009our}
\bibfield{author}{\bibinfo{person}{Mark~A Lemley}.}
  \bibinfo{year}{2009}\natexlab{}.
\newblock \showarticletitle{Our Bizarre System for Proving Copyright
  Infringement}.
\newblock \bibinfo{journal}{\emph{J. Copyright Soc'y USA}}
  \bibinfo{volume}{57} (\bibinfo{year}{2009}), \bibinfo{pages}{719}.
\newblock


\bibitem[Leong et~al\mbox{.}(2006)]%
        {leong2006randomness}
\bibfield{author}{\bibinfo{person}{Tuck~Wah Leong}, \bibinfo{person}{Frank
  Vetere}, {and} \bibinfo{person}{Steve Howard}.}
  \bibinfo{year}{2006}\natexlab{}.
\newblock \showarticletitle{Randomness as a resource for design}. In
  \bibinfo{booktitle}{\emph{Proceedings of the 6th conference on Designing
  Interactive systems}}. \bibinfo{pages}{132--139}.
\newblock


\bibitem[Lester(2013)]%
        {lester2013blurred}
\bibfield{author}{\bibinfo{person}{Toni Lester}.}
  \bibinfo{year}{2013}\natexlab{}.
\newblock \showarticletitle{Blurred Lines-Where Copyright Ends and Cultural
  Appropriation Begins-The Case of Robin Thicke versus Bridgeport Music and the
  Estate of Marvin Gaye}.
\newblock \bibinfo{journal}{\emph{Hastings Comm. \& Ent. LJ}}
  \bibinfo{volume}{36} (\bibinfo{year}{2013}), \bibinfo{pages}{217}.
\newblock


\bibitem[LeWitt(1967)]%
        {lewitt1967paragraphs}
\bibfield{author}{\bibinfo{person}{Sol LeWitt}.}
  \bibinfo{year}{1967}\natexlab{}.
\newblock \showarticletitle{Paragraphs on conceptual art}.
\newblock \bibinfo{journal}{\emph{Artforum}} \bibinfo{volume}{5},
  \bibinfo{number}{10} (\bibinfo{year}{1967}), \bibinfo{pages}{79--83}.
\newblock


\bibitem[Liu and Chilton(2022)]%
        {liu2022design}
\bibfield{author}{\bibinfo{person}{Vivian Liu} {and} \bibinfo{person}{Lydia~B
  Chilton}.} \bibinfo{year}{2022}\natexlab{}.
\newblock \showarticletitle{Design guidelines for prompt engineering
  text-to-image generative models}. In \bibinfo{booktitle}{\emph{Proceedings of
  the 2022 CHI Conference on Human Factors in Computing Systems}}.
  \bibinfo{pages}{1--23}.
\newblock


\bibitem[Lund(2011)]%
        {lund2011empirical}
\bibfield{author}{\bibinfo{person}{Jamie Lund}.}
  \bibinfo{year}{2011}\natexlab{}.
\newblock \showarticletitle{An empirical examination of the lay listener test
  in music composition copyright infringement}.
\newblock \bibinfo{journal}{\emph{Va. Sports \& Ent. LJ}}  \bibinfo{volume}{11}
  (\bibinfo{year}{2011}), \bibinfo{pages}{137}.
\newblock


\bibitem[Mackenzie(2017)]%
        {mackenzie2017machine}
\bibfield{author}{\bibinfo{person}{Adrian Mackenzie}.}
  \bibinfo{year}{2017}\natexlab{}.
\newblock \bibinfo{booktitle}{\emph{Machine learners: Archaeology of a data
  practice}}.
\newblock \bibinfo{publisher}{MIT Press}.
\newblock


\bibitem[Marks and Baldry(2009)]%
        {marks2009stuck}
\bibfield{author}{\bibinfo{person}{Abigail Marks} {and} \bibinfo{person}{Chris
  Baldry}.} \bibinfo{year}{2009}\natexlab{}.
\newblock \showarticletitle{Stuck in the middle with who? The class identity of
  knowledge workers}.
\newblock \bibinfo{journal}{\emph{Work, Employment and Society}}
  \bibinfo{volume}{23}, \bibinfo{number}{1} (\bibinfo{year}{2009}),
  \bibinfo{pages}{49--65}.
\newblock


\bibitem[Masad(2016)]%
        {masad_2016}
\bibfield{author}{\bibinfo{person}{Ilana Masad}.}
  \bibinfo{year}{2016}\natexlab{}.
\newblock \bibinfo{title}{Harry Potter and the possible queerbaiting: Why fans
  are mad over a lack of gay romance}.
\newblock
\newblock
\urldef\tempurl%
\url{https://www.theguardian.com/books/booksblog/2016/aug/16/harry-potter-possible-example-queerbaiting-cursed-child}
\showURL{%
\tempurl}


\bibitem[McLuhan(1994)]%
        {mcluhan1994understanding}
\bibfield{author}{\bibinfo{person}{Marshall McLuhan}.}
  \bibinfo{year}{1994}\natexlab{}.
\newblock \bibinfo{booktitle}{\emph{Understanding media: The extensions of
  man}}.
\newblock \bibinfo{publisher}{MIT press}.
\newblock


\bibitem[McRobbie(2013)]%
        {mcrobbie_2013}
\bibfield{author}{\bibinfo{person}{Linda~Rodriguez McRobbie}.}
  \bibinfo{year}{2013}\natexlab{}.
\newblock \bibinfo{title}{The strange and mysterious history of the Ouija
  board}.
\newblock
\newblock
\urldef\tempurl%
\url{https://www.smithsonianmag.com/history/the-strange-and-mysterious-history-of-the-ouija-board-5860627/}
\showURL{%
\tempurl}


\bibitem[Miller(2019a)]%
        {miller2019artist}
\bibfield{author}{\bibinfo{person}{Arthur~I Miller}.}
  \bibinfo{year}{2019}\natexlab{a}.
\newblock \bibinfo{booktitle}{\emph{The artist in the machine: The world of
  AI-powered creativity}}.
\newblock \bibinfo{publisher}{Mit Press}.
\newblock


\bibitem[Miller(2019b)]%
        {miller2019explanation}
\bibfield{author}{\bibinfo{person}{Tim Miller}.}
  \bibinfo{year}{2019}\natexlab{b}.
\newblock \showarticletitle{Explanation in artificial intelligence: Insights
  from the social sciences}.
\newblock \bibinfo{journal}{\emph{Artificial intelligence}}
  \bibinfo{volume}{267} (\bibinfo{year}{2019}), \bibinfo{pages}{1--38}.
\newblock


\bibitem[Mirowski et~al\mbox{.}(2022)]%
        {mirowski2022co}
\bibfield{author}{\bibinfo{person}{Piotr Mirowski}, \bibinfo{person}{Kory~W
  Mathewson}, \bibinfo{person}{Jaylen Pittman}, {and} \bibinfo{person}{Richard
  Evans}.} \bibinfo{year}{2022}\natexlab{}.
\newblock \showarticletitle{Co-writing screenplays and theatre scripts with
  language models: An evaluation by industry professionals}.
\newblock \bibinfo{journal}{\emph{arXiv preprint arXiv:2209.14958}}
  (\bibinfo{year}{2022}).
\newblock


\bibitem[Moffat(2004)]%
        {moffat2004mutant}
\bibfield{author}{\bibinfo{person}{Viva~R Moffat}.}
  \bibinfo{year}{2004}\natexlab{}.
\newblock \showarticletitle{Mutant copyrights and backdoor patents: the problem
  of overlapping intellectual property protection}.
\newblock \bibinfo{journal}{\emph{Berkeley Tech. LJ}}  \bibinfo{volume}{19}
  (\bibinfo{year}{2004}), \bibinfo{pages}{1473}.
\newblock


\bibitem[Moulier-Boutang(2011)]%
        {moulier2011cognitive}
\bibfield{author}{\bibinfo{person}{Yann Moulier-Boutang}.}
  \bibinfo{year}{2011}\natexlab{}.
\newblock \bibinfo{booktitle}{\emph{Cognitive capitalism}}.
\newblock \bibinfo{publisher}{Polity}.
\newblock


\bibitem[M{\"u}hl(1930)]%
        {muhl1930automatic}
\bibfield{author}{\bibinfo{person}{Anita~Mary M{\"u}hl}.}
  \bibinfo{year}{1930}\natexlab{}.
\newblock \bibinfo{booktitle}{\emph{Automatic writing}}.
\newblock \bibinfo{publisher}{T. Steinkopff}.
\newblock


\bibitem[Murphy(1990)]%
        {murphy1990jazz}
\bibfield{author}{\bibinfo{person}{John~P Murphy}.}
  \bibinfo{year}{1990}\natexlab{}.
\newblock \showarticletitle{Jazz improvisation: The joy of influence}.
\newblock \bibinfo{journal}{\emph{The black perspective in music}}
  (\bibinfo{year}{1990}), \bibinfo{pages}{7--19}.
\newblock


\bibitem[Musto(2010)]%
        {musto2010revisiting}
\bibfield{author}{\bibinfo{person}{Marcello Musto}.}
  \bibinfo{year}{2010}\natexlab{}.
\newblock \showarticletitle{Revisiting Marx's concept of alienation}.
\newblock \bibinfo{journal}{\emph{Socialism and Democracy}}
  \bibinfo{volume}{24}, \bibinfo{number}{3} (\bibinfo{year}{2010}),
  \bibinfo{pages}{79--101}.
\newblock


\bibitem[Musto(2015)]%
        {musto2015young}
\bibfield{author}{\bibinfo{person}{Marcello Musto}.}
  \bibinfo{year}{2015}\natexlab{}.
\newblock \showarticletitle{The ‘Young Marx’Myth in interpretations of the
  economic--philosophic manuscripts of 1844}.
\newblock \bibinfo{journal}{\emph{Critique}} \bibinfo{volume}{43},
  \bibinfo{number}{2} (\bibinfo{year}{2015}), \bibinfo{pages}{233--260}.
\newblock


\bibitem[Neves et~al\mbox{.}(2021)]%
        {neves2021link}
\bibfield{author}{\bibinfo{person}{Pedro~Cunha Neves}, \bibinfo{person}{Oscar
  Afonso}, \bibinfo{person}{Diana Silva}, {and} \bibinfo{person}{Elena
  Sochirca}.} \bibinfo{year}{2021}\natexlab{}.
\newblock \showarticletitle{The link between intellectual property rights,
  innovation, and growth: A meta-analysis}.
\newblock \bibinfo{journal}{\emph{Economic Modelling}}  \bibinfo{volume}{97}
  (\bibinfo{year}{2021}), \bibinfo{pages}{196--209}.
\newblock


\bibitem[Newman(1999)]%
        {newman1999new}
\bibfield{author}{\bibinfo{person}{Jon~O Newman}.}
  \bibinfo{year}{1999}\natexlab{}.
\newblock \showarticletitle{New lyrics for an old melody: The idea/expression
  dichotomy in the computer age}.
\newblock \bibinfo{journal}{\emph{Cardozo Arts \& Ent. LJ}}
  \bibinfo{volume}{17} (\bibinfo{year}{1999}), \bibinfo{pages}{691}.
\newblock


\bibitem[Nuttall(1968)]%
        {nuttall1968did}
\bibfield{author}{\bibinfo{person}{Anthony~David Nuttall}.}
  \bibinfo{year}{1968}\natexlab{}.
\newblock \showarticletitle{Did Meursault mean to kill the Arab?-The
  Intentional Fallacy Fallacy}.
\newblock \bibinfo{journal}{\emph{Critical Quarterly}} \bibinfo{volume}{10},
  \bibinfo{number}{1-2} (\bibinfo{year}{1968}), \bibinfo{pages}{95--106}.
\newblock


\bibitem[Oppenlaender(2022)]%
        {oppenlaender2022creativity}
\bibfield{author}{\bibinfo{person}{Jonas Oppenlaender}.}
  \bibinfo{year}{2022}\natexlab{}.
\newblock \showarticletitle{The Creativity of Text-to-Image Generation}. In
  \bibinfo{booktitle}{\emph{Proceedings of the 25th International Academic
  Mindtrek Conference}}. \bibinfo{pages}{192--202}.
\newblock


\bibitem[Palmer(2015)]%
        {palmer2015blurred}
\bibfield{author}{\bibinfo{person}{Jason Palmer}.}
  \bibinfo{year}{2015}\natexlab{}.
\newblock \showarticletitle{Blurred Lines Means Changing Focus: Juries Composed
  of Musical Artists Should Decide Music Copyright Infringement Cases, Not Lay
  Juries}.
\newblock \bibinfo{journal}{\emph{Vand. J. Ent. \& Tech. L.}}
  \bibinfo{volume}{18} (\bibinfo{year}{2015}), \bibinfo{pages}{907}.
\newblock


\bibitem[Pasquinelli(2009)]%
        {pasquinelli2009google}
\bibfield{author}{\bibinfo{person}{Matteo Pasquinelli}.}
  \bibinfo{year}{2009}\natexlab{}.
\newblock \showarticletitle{Google’s PageRank algorithm: A diagram of
  cognitive capitalism and the rentier of the common intellect}.
\newblock \bibinfo{journal}{\emph{Deep search: The politics of search beyond
  Google}} (\bibinfo{year}{2009}), \bibinfo{pages}{152--162}.
\newblock


\bibitem[Patterson(1965)]%
        {patterson1965statute}
\bibfield{author}{\bibinfo{person}{Lyman~Ray Patterson}.}
  \bibinfo{year}{1965}\natexlab{}.
\newblock \showarticletitle{The Statute of Anne: Copyright Misconstrued}.
\newblock \bibinfo{journal}{\emph{Harv. J. on Legis.}}  \bibinfo{volume}{3}
  (\bibinfo{year}{1965}), \bibinfo{pages}{223}.
\newblock


\bibitem[Perloff and Dworkin(2009)]%
        {perloff2009sound}
\bibfield{author}{\bibinfo{person}{Marjorie Perloff} {and}
  \bibinfo{person}{Craig Dworkin}.} \bibinfo{year}{2009}\natexlab{}.
\newblock \bibinfo{booktitle}{\emph{The Sound of Poetry/The Poetry of Sound}}.
\newblock \bibinfo{publisher}{University of Chicago Press}.
\newblock


\bibitem[Phillips et~al\mbox{.}(1987)]%
        {phillips1987humument}
\bibfield{author}{\bibinfo{person}{Tom Phillips} {et~al\mbox{.}}}
  \bibinfo{year}{1987}\natexlab{}.
\newblock \showarticletitle{A humument}.
\newblock  (\bibinfo{year}{1987}).
\newblock


\bibitem[Ploin et~al\mbox{.}(2022)]%
        {Ploin2022}
\bibfield{author}{\bibinfo{person}{A. Ploin}, \bibinfo{person}{R. Eynon},
  \bibinfo{person}{Hjorth I.}, {and} \bibinfo{person}{M.A. Osborne}.}
  \bibinfo{year}{2022}\natexlab{}.
\newblock \bibinfo{booktitle}{\emph{AI and the Arts: How Machine Learning is
  Changing Artistic Work}}.
\newblock \bibinfo{type}{Report from the Creative Algorithmic Intelligence
  Research Project}. \bibinfo{institution}{Oxford Internet Institute},
  \bibinfo{address}{University of Oxford, UK}.
\newblock


\bibitem[Pulvirent(2015)]%
        {pulvirent_2015}
\bibfield{author}{\bibinfo{person}{Stephen Pulvirent}.}
  \bibinfo{year}{2015}\natexlab{}.
\newblock \bibinfo{title}{How Daniel Wellington made a \$200 million business
  out of cheap watches}.
\newblock
\newblock
\urldef\tempurl%
\url{https://www.bloomberg.com/news/articles/2015-07-14/how-daniel-wellington-made-a-200-million-business-out-of-cheap-watches}
\showURL{%
\tempurl}


\bibitem[Ramesh et~al\mbox{.}(2021)]%
        {ramesh2021zero}
\bibfield{author}{\bibinfo{person}{Aditya Ramesh}, \bibinfo{person}{Mikhail
  Pavlov}, \bibinfo{person}{Gabriel Goh}, \bibinfo{person}{Scott Gray},
  \bibinfo{person}{Chelsea Voss}, \bibinfo{person}{Alec Radford},
  \bibinfo{person}{Mark Chen}, {and} \bibinfo{person}{Ilya Sutskever}.}
  \bibinfo{year}{2021}\natexlab{}.
\newblock \showarticletitle{Zero-shot text-to-image generation}. In
  \bibinfo{booktitle}{\emph{International Conference on Machine Learning}}.
  PMLR, \bibinfo{pages}{8821--8831}.
\newblock


\bibitem[Riehl(2020)]%
        {riehl_2020}
\bibfield{author}{\bibinfo{person}{Damien Riehl}.}
  \bibinfo{year}{2020}\natexlab{}.
\newblock \bibinfo{title}{All the Music}.
\newblock
\newblock
\urldef\tempurl%
\url{http://allthemusic.info/faqs/}
\showURL{%
\tempurl}


\bibitem[Rosati(2017)]%
        {rosati2017monkey}
\bibfield{author}{\bibinfo{person}{Eleonora Rosati}.}
  \bibinfo{year}{2017}\natexlab{}.
\newblock \showarticletitle{The Monkey Selfie case and the concept of
  authorship: an EU perspective}.
\newblock \bibinfo{journal}{\emph{Journal of Intellectual Property Law \&
  Practice}} \bibinfo{volume}{12}, \bibinfo{number}{12} (\bibinfo{year}{2017}),
  \bibinfo{pages}{973--977}.
\newblock


\bibitem[Rudolph et~al\mbox{.}(2023)]%
        {rudolph2023chatgpt}
\bibfield{author}{\bibinfo{person}{J{\"u}rgen Rudolph}, \bibinfo{person}{Samson
  Tan}, {and} \bibinfo{person}{Shannon Tan}.} \bibinfo{year}{2023}\natexlab{}.
\newblock \showarticletitle{ChatGPT: Bullshit spewer or the end of traditional
  assessments in higher education?}
\newblock \bibinfo{journal}{\emph{Journal of Applied Learning and Teaching}}
  \bibinfo{volume}{6}, \bibinfo{number}{1} (\bibinfo{year}{2023}).
\newblock


\bibitem[Sarkar(2023)]%
        {sarkar2023enough}
\bibfield{author}{\bibinfo{person}{Advait Sarkar}.}
  \bibinfo{year}{2023}\natexlab{}.
\newblock \showarticletitle{Enough With "Human-AI Collaboration"}. In
  \bibinfo{booktitle}{\emph{Extended Abstracts of the 2023 CHI Conference on
  Human Factors in Computing Systems (CHI EA 2023)}}.
\newblock


\bibitem[Sarkar et~al\mbox{.}(2022)]%
        {sarkar2022programmingai}
\bibfield{author}{\bibinfo{person}{Advait Sarkar}, \bibinfo{person}{Andrew~D.
  Gordon}, \bibinfo{person}{Carina Negreanu}, \bibinfo{person}{Christian
  Poelitz}, \bibinfo{person}{Sruti Srinivasa~Ragavan}, {and}
  \bibinfo{person}{Ben Zorn}.} \bibinfo{year}{2022}\natexlab{}.
\newblock \showarticletitle{What is it like to program with artificial
  intelligence?}. In \bibinfo{booktitle}{\emph{{Proceedings of the 33rd Annual
  Conference of the Psychology of Programming Interest Group (PPIG 2022)}}}.
\newblock


\bibitem[Schloss(2014)]%
        {schloss2014making}
\bibfield{author}{\bibinfo{person}{Joseph~G Schloss}.}
  \bibinfo{year}{2014}\natexlab{}.
\newblock \bibinfo{booktitle}{\emph{Making beats: The art of sample-based
  hip-hop}}.
\newblock \bibinfo{publisher}{Wesleyan University Press}.
\newblock


\bibitem[Sedgewick(2021)]%
        {sedgewick2021coffeeland}
\bibfield{author}{\bibinfo{person}{Augustine Sedgewick}.}
  \bibinfo{year}{2021}\natexlab{}.
\newblock \bibinfo{booktitle}{\emph{Coffeeland: One Man's Dark Empire and the
  Making of Our Favorite Drug}}.
\newblock \bibinfo{publisher}{Penguin}.
\newblock


\bibitem[Shaviro(2008)]%
        {shaviro_2008}
\bibfield{author}{\bibinfo{person}{Steven Shaviro}.}
  \bibinfo{year}{2008}\natexlab{}.
\newblock \bibinfo{title}{Cognitive capitalism?}
\newblock
\newblock
\urldef\tempurl%
\url{http://www.shaviro.com/Blog/?p=620}
\showURL{%
\tempurl}


\bibitem[Simonton(2012)]%
        {simonton2012taking}
\bibfield{author}{\bibinfo{person}{Dean~Keith Simonton}.}
  \bibinfo{year}{2012}\natexlab{}.
\newblock \showarticletitle{Taking the US Patent Office criteria seriously: A
  quantitative three-criterion creativity definition and its implications}.
\newblock \bibinfo{journal}{\emph{Creativity research journal}}
  \bibinfo{volume}{24}, \bibinfo{number}{2-3} (\bibinfo{year}{2012}),
  \bibinfo{pages}{97--106}.
\newblock


\bibitem[Singh et~al\mbox{.}(2022)]%
        {singh2022hide}
\bibfield{author}{\bibinfo{person}{Nikhil Singh}, \bibinfo{person}{Guillermo
  Bernal}, \bibinfo{person}{Daria Savchenko}, {and} \bibinfo{person}{Elena~L
  Glassman}.} \bibinfo{year}{2022}\natexlab{}.
\newblock \showarticletitle{Where to hide a stolen elephant: Leaps in creative
  writing with multimodal machine intelligence}.
\newblock \bibinfo{journal}{\emph{ACM Transactions on Computer-Human
  Interaction}} (\bibinfo{year}{2022}).
\newblock


\bibitem[Sisario(2019)]%
        {sisario_2019}
\bibfield{author}{\bibinfo{person}{Ben Sisario}.}
  \bibinfo{year}{2019}\natexlab{}.
\newblock \bibinfo{title}{'blurred lines' on their minds, songwriters create
  nervously}.
\newblock
\newblock
\urldef\tempurl%
\url{https://www.nytimes.com/2019/03/31/business/media/plagiarism-music-songwriters.html}
\showURL{%
\tempurl}


\bibitem[Smith(2021)]%
        {smith2021decolonizing}
\bibfield{author}{\bibinfo{person}{Linda~Tuhiwai Smith}.}
  \bibinfo{year}{2021}\natexlab{}.
\newblock \bibinfo{booktitle}{\emph{Decolonizing methodologies: Research and
  indigenous peoples}}.
\newblock \bibinfo{publisher}{Bloomsbury Publishing}.
\newblock


\bibitem[Smith(2022)]%
        {smith_2022}
\bibfield{author}{\bibinfo{person}{Noah Smith}.}
  \bibinfo{year}{2022}\natexlab{}.
\newblock \bibinfo{title}{Generative AI: Autocomplete for everything}.
\newblock
\newblock
\urldef\tempurl%
\url{https://noahpinion.substack.com/p/generative-ai-autocomplete-for-everything}
\showURL{%
\tempurl}


\bibitem[Smith(2006)]%
        {smith2006inkwell}
\bibfield{author}{\bibinfo{person}{Stephen Smith}.}
  \bibinfo{year}{2006}\natexlab{}.
\newblock \bibinfo{booktitle}{\emph{An inkwell of pen names}}.
\newblock \bibinfo{publisher}{Xlibris Corporation}.
\newblock


\bibitem[Snyder(2011)]%
        {snyder2011philosophical}
\bibfield{author}{\bibinfo{person}{Laura~J Snyder}.}
  \bibinfo{year}{2011}\natexlab{}.
\newblock \bibinfo{booktitle}{\emph{The philosophical breakfast club: four
  remarkable friends who transformed science and changed the world}}.
\newblock \bibinfo{publisher}{Crown}.
\newblock


\bibitem[Sontag et~al\mbox{.}(1994)]%
        {sontag1994against}
\bibfield{author}{\bibinfo{person}{Susan Sontag} {et~al\mbox{.}}}
  \bibinfo{year}{1994}\natexlab{}.
\newblock \bibinfo{booktitle}{\emph{Against interpretation}}.
\newblock \bibinfo{publisher}{Vintage London}.
\newblock


\bibitem[Suh et~al\mbox{.}(2021)]%
        {suh2021ai}
\bibfield{author}{\bibinfo{person}{Minhyang Suh}, \bibinfo{person}{Emily
  Youngblom}, \bibinfo{person}{Michael Terry}, {and} \bibinfo{person}{Carrie~J
  Cai}.} \bibinfo{year}{2021}\natexlab{}.
\newblock \showarticletitle{Ai as social glue: Uncovering the roles of deep
  generative ai during social music composition}. In
  \bibinfo{booktitle}{\emph{Proceedings of the 2021 CHI conference on human
  factors in computing systems}}. \bibinfo{pages}{1--11}.
\newblock


\bibitem[Thorp(2023)]%
        {doi:10.1126/science.adg7879}
\bibfield{author}{\bibinfo{person}{H.~Holden Thorp}.}
  \bibinfo{year}{2023}\natexlab{}.
\newblock \showarticletitle{ChatGPT is fun, but not an author}.
\newblock \bibinfo{journal}{\emph{Science}} \bibinfo{volume}{379},
  \bibinfo{number}{6630} (\bibinfo{year}{2023}), \bibinfo{pages}{313--313}.
\newblock
\urldef\tempurl%
\url{https://doi.org/10.1126/science.adg7879}
\showDOI{\tempurl}
\showeprint{https://www.science.org/doi/pdf/10.1126/science.adg7879}


\bibitem[van Heerden and Bas(2021)]%
        {van2021ai}
\bibfield{author}{\bibinfo{person}{Imke van Heerden} {and}
  \bibinfo{person}{Anil Bas}.} \bibinfo{year}{2021}\natexlab{}.
\newblock \showarticletitle{Ai as author--bridging the gap between machine
  learning and literary theory}.
\newblock \bibinfo{journal}{\emph{Journal of Artificial Intelligence Research}}
   \bibinfo{volume}{71} (\bibinfo{year}{2021}), \bibinfo{pages}{175--189}.
\newblock


\bibitem[VanEenoo(2011)]%
        {vaneenoo2011minimalism}
\bibfield{author}{\bibinfo{person}{Cedric VanEenoo}.}
  \bibinfo{year}{2011}\natexlab{}.
\newblock \showarticletitle{Minimalism in Art and Design: Concept, influences,
  implications and perspectives}.
\newblock \bibinfo{journal}{\emph{Journal of Fine and Studio Art}}
  \bibinfo{volume}{2}, \bibinfo{number}{1} (\bibinfo{year}{2011}),
  \bibinfo{pages}{7--12}.
\newblock


\bibitem[Vezzoso(2012)]%
        {vezzoso2012copyright}
\bibfield{author}{\bibinfo{person}{Simonetta Vezzoso}.}
  \bibinfo{year}{2012}\natexlab{}.
\newblock \showarticletitle{Copyright, Interfaces, and a Possible Atlantic
  Divide}.
\newblock \bibinfo{journal}{\emph{J. Intell. Prop. Info. Tech. \& Elec. Com.
  L.}}  \bibinfo{volume}{3} (\bibinfo{year}{2012}), \bibinfo{pages}{153}.
\newblock


\bibitem[Wallas(1926)]%
        {wallas1926art}
\bibfield{author}{\bibinfo{person}{Graham Wallas}.}
  \bibinfo{year}{1926}\natexlab{}.
\newblock \bibinfo{booktitle}{\emph{The art of thought}}.
  Vol.~\bibinfo{volume}{10}.
\newblock \bibinfo{publisher}{Harcourt, Brace}.
\newblock


\bibitem[Wang and Nickerson(2017)]%
        {wang2017literature}
\bibfield{author}{\bibinfo{person}{Kai Wang} {and} \bibinfo{person}{Jeffrey~V
  Nickerson}.} \bibinfo{year}{2017}\natexlab{}.
\newblock \showarticletitle{A literature review on individual creativity
  support systems}.
\newblock \bibinfo{journal}{\emph{Computers in Human Behavior}}
  \bibinfo{volume}{74} (\bibinfo{year}{2017}), \bibinfo{pages}{139--151}.
\newblock


\bibitem[Wang(2022)]%
        {wang_2022}
\bibfield{author}{\bibinfo{person}{Vivian Wang}.}
  \bibinfo{year}{2022}\natexlab{}.
\newblock \bibinfo{title}{A protest? A vigil? in Beijing, anxious crowds are
  unsure how far to go.}
\newblock
\newblock
\urldef\tempurl%
\url{https://www.nytimes.com/2022/11/28/world/asia/china-protests-covid-beijing.html}
\showURL{%
\tempurl}


\bibitem[Weidinger et~al\mbox{.}(2021)]%
        {weidinger2021ethical}
\bibfield{author}{\bibinfo{person}{Laura Weidinger}, \bibinfo{person}{John
  Mellor}, \bibinfo{person}{Maribeth Rauh}, \bibinfo{person}{Conor Griffin},
  \bibinfo{person}{Jonathan Uesato}, \bibinfo{person}{Po-Sen Huang},
  \bibinfo{person}{Myra Cheng}, \bibinfo{person}{Mia Glaese},
  \bibinfo{person}{Borja Balle}, \bibinfo{person}{Atoosa Kasirzadeh},
  {et~al\mbox{.}}} \bibinfo{year}{2021}\natexlab{}.
\newblock \showarticletitle{Ethical and social risks of harm from language
  models}.
\newblock \bibinfo{journal}{\emph{arXiv preprint arXiv:2112.04359}}
  (\bibinfo{year}{2021}).
\newblock


\bibitem[Weil(2023)]%
        {weil_2023}
\bibfield{author}{\bibinfo{person}{Elizabeth Weil}.}
  \bibinfo{year}{2023}\natexlab{}.
\newblock \bibinfo{title}{You are not a parrot}.
\newblock
\newblock
\urldef\tempurl%
\url{https://nymag.com/intelligencer/article/ai-artificial-intelligence-chatbots-emily-m-bender.html}
\showURL{%
\tempurl}


\bibitem[Wetzel(2018)]%
        {sep-types-tokens}
\bibfield{author}{\bibinfo{person}{Linda Wetzel}.}
  \bibinfo{year}{2018}\natexlab{}.
\newblock \showarticletitle{{Types and Tokens}}.
\newblock In \bibinfo{booktitle}{\emph{The {Stanford} Encyclopedia of
  Philosophy} (\bibinfo{edition}{{F}all 2018} ed.)},
  \bibfield{editor}{\bibinfo{person}{Edward~N. Zalta}} (Ed.).
  \bibinfo{publisher}{Metaphysics Research Lab, Stanford University}.
\newblock


\bibitem[Wimsatt and Beardsley(1946)]%
        {wimsatt1946intentional}
\bibfield{author}{\bibinfo{person}{William~Kurtz Wimsatt} {and}
  \bibinfo{person}{Monroe~Curtis Beardsley}.} \bibinfo{year}{1946}\natexlab{}.
\newblock \showarticletitle{The intentional fallacy}.
\newblock \bibinfo{journal}{\emph{The sewanee review}} \bibinfo{volume}{54},
  \bibinfo{number}{3} (\bibinfo{year}{1946}), \bibinfo{pages}{468--488}.
\newblock


\bibitem[Woods and Ma(2023)]%
        {woods_ma_2023}
\bibfield{author}{\bibinfo{person}{Darian Woods} {and} \bibinfo{person}{Adrian
  Ma}.} \bibinfo{year}{2023}\natexlab{}.
\newblock \bibinfo{title}{Artists vs. AI}.
\newblock
\newblock
\urldef\tempurl%
\url{https://www.npr.org/transcripts/1152653269}
\showURL{%
\tempurl}


\bibitem[Xu et~al\mbox{.}(2023)]%
        {xu2023creativity}
\bibfield{author}{\bibinfo{person}{Tianna Xu}, \bibinfo{person}{Advait Sarkar},
  {and} \bibinfo{person}{Sean Rintel}.} \bibinfo{year}{2023}\natexlab{}.
\newblock \showarticletitle{Is a Return To Office a Return To Creativity?
  Requiring Fixed Time In Office To Enable Brainstorms and Watercooler Talk May
  Not Foster Research Creativity} \emph{(\bibinfo{series}{CHIWORK 2023})}.
  \bibinfo{publisher}{Association for Computing Machinery},
  \bibinfo{address}{New York, NY, USA}.
\newblock


\bibitem[Yang et~al\mbox{.}(2022)]%
        {yang2022ai}
\bibfield{author}{\bibinfo{person}{Daijin Yang}, \bibinfo{person}{Yanpeng
  Zhou}, \bibinfo{person}{Zhiyuan Zhang}, \bibinfo{person}{Toby Jia-Jun Li},
  {and} \bibinfo{person}{Ray LC}.} \bibinfo{year}{2022}\natexlab{}.
\newblock \showarticletitle{AI as an Active Writer: Interaction strategies with
  generated text in human-AI collaborative fiction writing}. In
  \bibinfo{booktitle}{\emph{Joint Proceedings of the ACM IUI Workshops}},
  Vol.~\bibinfo{volume}{10}.
\newblock


\bibitem[Yule and Widdowson(1996)]%
        {yule1996pragmatics}
\bibfield{author}{\bibinfo{person}{George Yule} {and} \bibinfo{person}{HG
  Widdowson}.} \bibinfo{year}{1996}\natexlab{}.
\newblock \bibinfo{booktitle}{\emph{Pragmatics}}.
\newblock \bibinfo{publisher}{Oxford university press}.
\newblock


\bibitem[Zimmerman et~al\mbox{.}(2007)]%
        {zimmerman2007research}
\bibfield{author}{\bibinfo{person}{John Zimmerman}, \bibinfo{person}{Jodi
  Forlizzi}, {and} \bibinfo{person}{Shelley Evenson}.}
  \bibinfo{year}{2007}\natexlab{}.
\newblock \showarticletitle{Research through design as a method for interaction
  design research in HCI}. In \bibinfo{booktitle}{\emph{Proceedings of the
  SIGCHI conference on Human factors in computing systems}}.
  \bibinfo{pages}{493--502}.
\newblock


\end{thebibliography}










\end{document}